\documentclass[12pt]{article}
\usepackage{graphics,epsfig, psfrag}
\def \HT{{\mathcal H}}

\begin{document}

\makeatletter
\@addtoreset{equation}{section}
\makeatother
\renewcommand{\theequation}{\thesection.\arabic{equation}}

\begin{titlepage}

\baselineskip =15.5pt
\pagestyle{plain}
\setcounter{page}{0}

\vfil

\begin{center}
{\huge Cartan Calculus via Pauli Matrices}
\end{center}

\vfil

\begin{center}
{\large D. Mauro}\\
\vspace {1mm}
Dipartimento di Fisica Teorica, Universit\`a di Trieste, \\
Strada Costiera 11, P.O.Box 586, Trieste, Italy \\ and INFN, Sezione 
di Trieste.\\
e-mail: mauro@ts.infn.it
\vspace {1mm}
\vspace{3mm}
\end{center}

\vfil

\noindent In this paper we will provide a {\it new} operatorial counterpart of the path-integral 
formalism of classical mechanics developed in recent years. We call it {\it new} because
the Jacobi fields and forms will be realized via finite dimensional matrices. As a byproduct 
of this we will prove 
that all the operations of the Cartan calculus, such as the exterior derivative, the interior contraction 
with a vector field, the
Lie derivative and so on, can be realized by means of suitable tensor products of Pauli and 
identity matrices. 

\noindent 

\vfil
\end{titlepage}
\newpage

\section{Introduction}
At the end of the Eighties in Ref. \cite{Gozzi} a path integral formulation for classical mechanics was put
forward: from now on we will indicate it with the acronym CPI for {\bf{C}}lassical 
{\bf{P}}ath {\bf{I}}ntegral. This formulation is the functional
counterpart of the operatorial version of classical mechanics developed in the Thirties by Koopman and
von Neumann \cite{Koopman}-\cite{von Neumann}. In the CPI, besides the standard
phase space, some auxiliary variables made their appearance. Both the physical \cite{Liapunov} 
and the geometrical \cite{Gozzi}-\cite{Mauro} meaning of these variables have been studied 
in detail. Some of these variables are Grassmannian, i.e. anticommuting c-numbers representing differential 
forms. Anyhow, as Coleman said in his Erice's lectures, \cite{Coleman} ``{\it 
anticommuting c-numbers are notoriously objects that make strong men quail}".
So the main goal of this paper is to replace the anticommuting variables with less mysterious
objects such as matrices. In performing this operation all the geometrical richness of the original CPI is not lost.
In fact in this paper we will show how it is possible to map the differential forms of a 
$2n$-dimensional phase space into $2^{2n}$-dimensional vectors. Correspondently all the standard operations
of differential geometry, such as exterior derivatives, Lie derivatives
and so on, can be translated in terms of suitable combinations of Pauli and identity matrices.

The paper is organized as follows:
in Section {\bf 2} we briefly review the functional formulation of classical mechanics. 
In Section {\bf 3} we show how, in the
case of one degree of freedom $n=1$, it is possible to represent the fermionic operators of the CPI 
in terms of square matrices. In Section {\bf 4} we write in our formalism the universally
conserved charges originally discovered in the CPI \cite{Gozzi} and link them 
to the above mentioned operations of differential geometry. Sections {\bf 5}
and {\bf 6} contain the generalization of the previous results to the case of a system 
with an arbitrary great number of degrees of freedom. 
In Appendix {\bf A} we show the manner to reconstruct the evolution of a generalized 
wave function at the matrix level. In Appendix {\bf B} we show that the 
symmetry charges of the CPI make a superalgebra whose irreducible representations can be constructed
as it is usually done in the literature \cite{Rittenberg}. 
In Appendices {\bf C} and {\bf D} we confine some further calculational details. 

\section{Path Integral for Classical Mechanics (CPI)}
In this section we want to review some of the basic points of the functional
formulation for classical mechanics. We will be very brief here because the interested reader can 
find a more detailed treatment in the original papers \cite{Gozzi}.
The basic idea of the CPI is to give a path integral 
formulation for classical mechanics.

Suppose we start with a dynamical system made of a $2n$-dimensional phase space ${\cal M}$ with 
coordinates $\varphi^a=(q^1,\ldots, q^n,p^1,\ldots, p^n)$, 
a Hamiltonian $H(\varphi)$ and the symplectic matrix 
$\omega^{ab}$. The 
Hamilton's equations of motion can then be written in the following compact form:
\begin{equation}
\displaystyle \dot{\varphi}^a=\omega^{ab}\partial_bH(\varphi). \label{hamilton}
\end{equation}
The probability of finding the system at a point $\varphi^a$ at time $t$, if it was at the point 
$\varphi_i^a$ at the initial time $t_i$,
is one if the point $\varphi^a$ is on the classical path $\varphi_{cl}^a(t)$ and zero otherwise
(with
classical path we mean the path that solves the equations of motion
with initial conditions $\varphi_i^a$).
The classical path integral that gives such a
probability can be easily expressed  by means of a functional Dirac delta forcing 
the system to stay on the classical path $\varphi_{cl}^a(t)$. So the transition probability
$K(\;\;|\;\;)$ can be written as
\begin{equation}
\displaystyle 
K(\varphi^a t|\varphi_i^a t_i)=N\int {\cal D}\varphi\;\widetilde{\delta}
[\varphi^a(t)-\varphi_{cl}^a(t)]. \label{ka}
\end{equation}
Following the steps illustrated in detail in Ref. \cite{Gozzi} it is possible to rewrite 
the functional Dirac delta in (\ref{ka})
as a Dirac delta on the equations of motion times a suitable functional determinant:
\begin{equation}
\widetilde{\delta}[\varphi^a-\varphi^a_{cl}]=
\widetilde{\delta}[\dot{\varphi}^a-\omega^{ab}\partial_bH]\textrm{det}[\delta_b^a\partial_t-\omega^{ac}
\partial_c\partial_bH].
\end{equation}
Exponentiating the functional Dirac delta by means of auxiliary variables $\lambda_a$ and the 
determinant 
by means of a couple of Grassmann variables $c^a,\overline{c}_a$, one gets the 
following expression
for the classical path integral:
\begin{equation}
\displaystyle 
K(\varphi^a t|\varphi^a_it_i) =N\int {\cal D}\varphi^a{\cal D}\lambda_a{\cal D}
c^a{\cal D}\overline{c}_a \;\textrm{exp}
\biggl[i\int dt{\cal L}\biggr] \label{CPI}
\end{equation}
where the Lagrangian is given by:
\begin{equation}
{\cal L}=\lambda_a\dot{\varphi}^a+i\overline{c}_a\dot{c}^a-\HT,\;\;\;\;\;\;\;\textrm{with}\;\;\;\;\;\;\HT=\lambda_a
\omega^{ab}\partial_bH
+i\overline{c}_a\omega^{ac}\partial_c\partial_bHc^b. \label{Lag}
\end{equation}
From the Lagrangian (\ref{Lag}) one can derive, besides the standard Hamilton's equations (\ref{hamilton})
for $\varphi$, the following equations of motion for the Grassmann variables:
\begin{eqnarray}
&&\dot{c}^b=\omega^{bc}\partial_c\partial_aHc^ a\nonumber\\
&&\dot{\overline{c}}_b=-\overline{c}_a\omega^{ac}\partial_c\partial_bH.
\end{eqnarray}
So infinitesimal transformations generated by ${\cal H}$ are given by:
\begin{eqnarray}
\displaystyle 
&&\varphi^{a^{\prime}}=\varphi^a+\epsilon\omega^{ab}\partial_bH\nonumber\\
&&c^{a^{\prime}}=c^a+\epsilon\omega^{ac}\partial_c\partial_bHc^b=
\frac{\partial\varphi^{a^{\prime}}}{\partial\varphi^b}c^b\nonumber\\
&&\overline{c}_a^{\prime}=\overline{c}_a
-\epsilon\overline{c}_b\omega^{bc}\partial_c\partial_aH=\frac{\partial\varphi^b}
{\partial\varphi^{a^{\prime}}}\overline{c}_b.
\end{eqnarray}
We notice from these equations that $c^a$ transforms, under the diffeomorphism generated by $\HT$, as a basis for 
the differential forms 
$d\varphi^a$, while $\overline{c}_a$ transforms as a basis for the vector fields 
$\displaystyle \frac{\partial}{\partial \varphi^a}$, see \cite{Gozzi} and \cite{Regini}.
From the kinetic part of the Lagrangian (\ref{Lag}) we can derive, as Feynman did for 
quantum mechanics,
the graded commutators of the theory. They are: 
\begin{equation}
\langle [\varphi^a, \lambda_b]_-\rangle=i\delta_b^a,\;\;\;\;\;\;\langle [\overline{c}_
b,c^a]_+\rangle=\delta_b^a.
\end{equation}
All other commutators are zero. In order to satisfy the above commutators we can realize
$\varphi^a$ and $c^a$ as multiplication operators and $\lambda_a$ and $\overline{c}_a$ as derivative ones:
\begin{equation}
\displaystyle \lambda_a=-i\frac{\partial}{\partial\varphi^a},\;\;\;\;\overline{c}_a=
\frac{\partial}{\partial c^a}.
\end{equation}
Substituting the previous relations into the Hamiltonian $\HT$ of Eq. (\ref{Lag}) we 
obtain the following operator (from now on we will omit the hat signs $\;\widehat{ }\;$ on the symbols of the
abstract operators. Instead we will use them on the symbols which indicate the matrices associated to
these operators):
\begin{equation}
\displaystyle {\cal H}=-i\omega^{ab}\partial_bH\partial_a-i\omega^{ac}\partial_c
\partial_bHc^b
\frac{\partial}{\partial c^a}. \label{fed}
\end{equation}
The first term of (\ref{fed}) is just the Liouville operator that appears in the operatorial 
formulation of classical
mechanics due to Koopman and von Neumann, \cite{Koopman}-\cite{von Neumann}. 
This confirms that (\ref{CPI}) is 
just the correct
functional counterpart of the operatorial formulation for classical mechanics. 
Moreover, thanks to the presence of Grassmann variables, the CPI 
provides also the evolution of 
more generalized objects \cite{Gozzi}. In fact the following kernel:
\begin{equation}
K(\varphi_f, c_f, t_f\arrowvert\varphi_i, c_i, t_i)=\int {\cal D}^{\prime\prime}
\varphi{\cal D}\lambda{\cal D}^{\prime\prime}c
{\cal D}\overline{c} \; \textrm{exp}\Biggl(i\int_{t_i}^{t_f}dt {\cal L}\Biggr)
\end{equation}
with the boundary conditions:
\begin{equation}
\varphi^{a}(t_i)=\varphi^{a}_i, \; \;\;\; \varphi^{a}(t_f)=\varphi^{a}_f, \;\;\;\;
c^{a}(t_i)=c^{a}_i, \;\;\;\; c^{a}(t_f)=c^{a}_f
\end{equation}
gives the evolution not only of the functions of $\varphi$ but also of the 
generalized wave functions
$\psi(\varphi,c)$, that live in the Hilbert space underlying the CPI \cite{dgm}. $\psi(\varphi,c)$
can be interpreted as the most general differential form on the symplectic space
${\cal M}$ \cite{Gozzi}\cite{Regini}.

Even if the auxiliary variables $\lambda_a,c^a, \overline{c}_a$ have all a well-defined geometrical
meaning \cite{Gozzi}\cite{Regini}\cite{dgm}, still the reader could claim that they are somehow
redundant because we can do classical mechanics by using just the $\varphi^a$ variables. This redundancy
is actually signaled by the presence of some universal symmetries \cite{Gozzi} whose charges
are:
\begin{eqnarray}
&&\displaystyle Q\equiv ic^a\lambda_a, \;\;\;\;\;\;\;\;\;
\overline{Q}\equiv  i\overline{c}_a\omega^{ab}\lambda_b,\nonumber\\
&&Q_f\equiv c^a\overline{c}_a,\;\;\;\;K\equiv \frac{1}{2}\omega_{ab}c^ac^b,\;\;\;\; 
\overline{K}\equiv \frac{1}{2}
\omega^{ab}\overline{c}_a\overline{c}_b.
\end{eqnarray}
There is also supersymmetry \cite{Gozzi}\cite{Deotto} whose charges are:
\begin{eqnarray}
&&Q_{\scriptscriptstyle H}=Q-\beta N=ic^a\lambda_a-\beta 
c^a\partial_aH\nonumber\\
&&\overline{Q}_{\scriptscriptstyle H}=\overline{Q}+
\beta\overline{N}=i\overline{c}_a\omega^{ab}\lambda_b+\beta\overline{c}_a\omega^{ab}
\partial_bH
\end{eqnarray}
and whose anticommutator gives $\HT$:
\begin{equation}
[Q_{\scriptscriptstyle H},\overline{Q}_{\scriptscriptstyle H}]_{\scriptscriptstyle +}
=2i\beta\HT.
\end{equation}
All these conserved charges play a very important role in the Cartan calculus as it 
is explained in detail
in \cite{Gozzi} and \cite{Mauro}.
In particular, since $c^a$ transforms as a basis for the forms 
and since there is a
natural mapping between the wedge product of forms and the multiplication of 
Grassmann
variables, it is possible to map \cite{Gozzi} every $p$-form into a function of the operators
$\varphi$ and $c$:
\begin{eqnarray}
&&F^{\scriptscriptstyle (p)}=F_{a_1\cdots a_p}(\varphi)d\varphi^{a_1}\wedge 
d\varphi^{a_2}\wedge \cdots
\wedge d\varphi^{a_p}\nonumber\\
&&\qquad \qquad \qquad \qquad \qquad \Updownarrow\nonumber\\
&&\qquad \qquad \widehat{F}^{\scriptscriptstyle (p)}=F_{a_1\cdots a_p}(\varphi)
c^{a_1}c^{a_2} \cdots c^{a_p}
\end{eqnarray}
In the same way, from the properties of transformation of $\overline{c}$ as a basis 
for vector fields it is possible 
to represent every vector field $V$ as a function of $\varphi$ and $\overline{c}$: 
$\widehat{V}
=V^ a(\varphi)\overline{c}_a$.
Via the commutators of the theory all the most important operations of the Cartan 
calculus can be reproduced \cite{Gozzi}.
In particular:\newline
$\bullet$ The $Q$ charge can be interpreted as the exterior
derivative acting on $p$-forms according to the relation:
\begin{equation}
{\bf d}F^{\scriptscriptstyle (p)} =[Q,\widehat{F}^
{\scriptscriptstyle (p)}].
\end{equation}
$\bullet$ The interior contraction of a $p$-form with a vector field $V$ can be reproduced 
by means of the following commutator:
\begin{equation}
\iota_{\scriptscriptstyle V}F^{\scriptscriptstyle (p)}=[\widehat{V},\widehat{F}^
{\scriptscriptstyle (p)}].
\end{equation}
$\bullet$ The Lie derivative along a Hamiltonian vector field \cite{Marsden}
$h^a=\omega^{ab}\partial_bH$ 
is given, modulus a factor $i$, by the operator of evolution $\HT$:
\begin{equation}
{\cal L}_hF^{\scriptscriptstyle (p)}=[i\HT, \widehat{F}^{\scriptscriptstyle (p)}] 
\end{equation}
and the well-known fact that the Lie derivative commutes with the exterior derivative ${\bf d}$: 
$[{\bf d},{\cal L}_h]=0$ implies immediately that $[Q,\HT]=0$.
For the geometrical interpretation of the supersymmetry charges we refer the interested
reader to Ref. \cite{Deotto}.\newline
In Sections {\bf 4} and {\bf 6} we will show how the previous operations of the Cartan calculus can be 
reproduced also via suitable combinations of Pauli and identity matrices.

\section{Grassmannian Operators and Matrices}
It is well-known
that every Grassmann algebra can be realized in terms of suitable square matrices, 
see for example the exercise (6.18) of Ref. \cite{Henneaux}.
This matrix realization was already used in supersymmetric
quantum mechanics \cite{Witten}\cite{Salomonson}. In this section we
want to see if the same matrix realization can be used for the CPI. 
In particular in this section we want to work out things in the case of a system 
with one degree of freedom $n=1$, i.e. $\varphi^a=(q,p),\;c^a=(c^q,c^p)$.
First of all  let us notice that the generalized wave function
\begin{equation}
\displaystyle 
\psi(\varphi,c)=\psi_{\scriptscriptstyle 0}(\varphi)+\psi_q(\varphi)c^q+\psi_p(\varphi)c^p+
\psi_{\scriptscriptstyle 2}(\varphi)c^pc^q  \label{genw}
\end{equation}
contains only 4 arbitrary functions $\psi_{\scriptscriptstyle 0}, \psi_q,\psi_p,\psi_{\scriptscriptstyle 2}$
and we could try to represent this object as a 4-vector made up of 4 bosonic components:
\begin{equation}
\displaystyle 
\psi(\varphi,c)=\psi_{\scriptscriptstyle 0}(\varphi)+\psi_q(\varphi)c^q+\psi_p(\varphi)c^p+
\psi_{\scriptscriptstyle 2}(\varphi)c^pc^q\equiv
\left( \begin{array}{c}
\psi_{\scriptscriptstyle 0}\\ \psi_q\\ \psi_p\\ \psi_{\scriptscriptstyle 2} \end{array}\right).  \label{genw2}
\end{equation}
With this choice it is then possible to represent every operator of the theory as a 
well-defined $4\times 4$ matrix. For example
if we take the operator of multiplication by $c^q$ and apply it on $\psi$ we get
$\psi^{\prime}=c^q\psi=c^q\psi_{\scriptscriptstyle 0}+c^qc^p\psi_p$.
This wave function $\psi^{\prime}$, in the 4-vector notation (\ref{genw2}), has the form:
\begin{equation}
\psi^{\prime}= \left( \begin{array}{c} 0\\ 
\psi_{\scriptscriptstyle 0}\\ 0 \\ 
-\psi_p \end{array}\right)
\end{equation}
and it could be obtained from the 4-vector representation of $\psi$ as
\begin{equation}
\psi^{\prime}=\pmatrix{0 & 0 & 0 & 0\cr
1 & 0 & 0 & 0\cr
0 & 0 & 0 & 0\cr
0 & 0 & -1 & 0} \left( \begin{array}{c}
\psi_{\scriptscriptstyle 0}\\ \psi_q\\ \psi_p\\ \psi_{\scriptscriptstyle 2} \end{array}\right).
\end{equation}
So we have the following matrix representation for the operator $c^q$:
\begin{equation}
\widehat{c}^q=\pmatrix{0 & 0 & 0 & 0\cr
1 & 0 & 0 & 0\cr
0 & 0 & 0 & 0\cr
0 & 0 & -1 & 0}. \label{operator1}
\end{equation}
In the same way the operator of multiplication by $c^p$ 
can be represented by the following $4\times 4$ matrix:
\begin{equation}
\widehat{c}^p=\pmatrix{0 & 0 & 0 & 0\cr
0 & 0 & 0 & 0\cr
1 & 0 & 0 & 0\cr
0 & 1 & 0 & 0}. \label{operator2}
\end{equation}
$\overline{c}_q$ and $\overline{c}_p$ are instead the derivative operators 
$\displaystyle \frac{\partial}{\partial c^q}$ and 
$\displaystyle \frac{\partial}{\partial c^p}$. If we apply them on the $\psi$ written as in (\ref{genw})
and we perform steps similar to those which lead to (\ref{operator1}), then we can obtain the following matrix 
representation:
\begin{equation}
\widehat{\overline{c}}_q=\pmatrix{0 & 1 & 0 & 0\cr
0 & 0 & 0 & 0\cr
0 & 0 & 0 & -1\cr
0 & 0 & 0 & 0}, \qquad\quad \widehat{\overline{c}}_p=\pmatrix{0 & 0 & 1 & 0\cr
0 & 0 & 0 & 1\cr
0 & 0 & 0 & 0\cr
0 & 0 & 0 & 0}. \label{operator3}
\end{equation}
Note that the matrices $\widehat{\overline{c}}$ are just the transpose 
of the associated matrices $\widehat{c}$.
It is also easy to verify that the matrices (\ref{operator1})-(\ref{operator3}) 
satisfy the correct anticommutation relations:
$[\widehat{c},\widehat{c}]_+=0$, $[\widehat{\overline{c}},\widehat{\overline{c}}]_+=0$, 
$[\widehat{c}^a,\widehat{\overline{c}}_b]_+=\delta_b^a$.

The $4\times 4$ matrices we have obtained so far can be written in a more compact form via Pauli
matrices. Let us introduce the matrices: 
\begin{equation}
\displaystyle
\frac{\sigma^{\scriptscriptstyle (+)}}{2}=\frac{\sigma_x+i\sigma_y}{2}=
\pmatrix{0 & 1\cr 0 & 0},\qquad\quad \frac{\sigma^{\scriptscriptstyle (-)}}{2}
=\frac{\sigma_x-i\sigma_y}{2}=
\pmatrix{0 & 0\cr 1 & 0}.
\end{equation}
It is then easy to prove that the matrices (\ref{operator1})-(\ref{operator3})
can be written as:
\begin{eqnarray}
\displaystyle
&&\widehat{c}^p=\frac{\sigma^{\scriptscriptstyle(-)}}{2}\otimes {\bf 1},\qquad\quad 
\widehat{c}^q=\sigma_z\otimes 
\frac{\sigma^{\scriptscriptstyle(-)}}{2}\nonumber\\
&&\widehat{\overline{c}}_p=\frac{\sigma^{\scriptscriptstyle(+)}}{2}\otimes {\bf 1},
\qquad\quad \widehat{\overline{c}}_q=\sigma_z\otimes 
\frac{\sigma^{\scriptscriptstyle(+)}}{2}
\end{eqnarray}
where $\otimes$ indicates the tensor product.
These formulas are very useful because, as we will see in Section {\bf 5}, 
they can be generalized to the case
of systems with an arbitrary great number of degrees of freedom.

Via the representation (\ref{operator1})-(\ref{operator3}) 
for the Grassmannian operators of the theory we can build also the
matrix representation of the symmetry charges present in the CPI:
\begin{eqnarray}
\displaystyle 
&&\widehat{Q}=i\widehat{c}^q\lambda_q+i\widehat{c}^p\lambda_p=\widehat{c}^q\partial_q+
\widehat{c}^p\partial_p=\pmatrix{0 & 0 & 0 & 0\cr \partial_q & 0 & 0 & 0\cr 
\partial_p & 0 & 0 & 0\cr 0 & \partial_p
& -\partial_q  & 0}, \nonumber\\
&& \widehat{\overline{Q}}=\pmatrix{0 & \partial_p & -\partial_q & 0\cr
0 & 0 & 0 & -\partial_q\cr 0 & 0 & 0 & -\partial_p\cr 0 & 0 & 0 & 0},
\quad \widehat{Q}_f=\pmatrix{0 & 0 & 0 & 0\cr 0 & 1 & 0 & 0\cr 0 & 0 & 1 & 0\cr 0 & 0
& 0  & 2}\nonumber\\
\bigskip
&&\widehat{K}=\pmatrix{0 & 0 & 0 & 0\cr 0 & 0 & 0 & 0\cr 0 & 0 & 0 & 0\cr 1 & 0
& 0 & 0}, \qquad \quad \widehat{\overline{K}}=\pmatrix{0 & 0 & 0 & 1\cr
0 & 0 & 0 & 0\cr 0 & 0 & 0 & 0\cr 0 & 0 & 0 & 0}. \label{opmat}
\end{eqnarray}
One can easily check that the algebra of these charges is the one of Ref. \cite{Gozzi}:
\begin{eqnarray}
&&[\widehat{Q},\widehat{Q}]_{\scriptscriptstyle +}=[\widehat{\overline{Q}},
\widehat{\overline{Q}}]_{\scriptscriptstyle +}=[\widehat{Q},\widehat{\overline{Q}}]_
{\scriptscriptstyle +}=0\nonumber\\
&&[\widehat{Q}_f,\widehat{K}]_{\scriptscriptstyle -}=2\widehat{K},\;\;[\widehat{Q}_f,
\widehat{\overline{K}}]_{\scriptscriptstyle -}=-2\widehat{\overline{K}},\;\;
[\widehat{K},\widehat{\overline{K}}]_{\scriptscriptstyle -}=\widehat{Q}_f-{\bf 1}\nonumber\\
&&[\widehat{Q}_f,\widehat{Q}]_{\scriptscriptstyle -}=\widehat{Q},\;\;
[\widehat{Q}_f,\widehat{\overline{Q}}]_{\scriptscriptstyle -}=
-\widehat{\overline{Q}},\;\; [\widehat{K},\widehat{Q}]_{\scriptscriptstyle -}=0\nonumber\\
&&[\widehat{K},\widehat{\overline{Q}}]_{\scriptscriptstyle -}=\widehat{Q},\;\;
[\widehat{\overline{K}},\widehat{Q}]_{\scriptscriptstyle -}=\widehat{\overline{Q}},\;\;
[\widehat{\overline{K}}, \widehat{\overline{Q}}]_{\scriptscriptstyle -}=0. \label{algcpi}
\end{eqnarray}
The representation of the supersymmetry charges is given by:
\begin{eqnarray}
\bigskip
&&\widehat{Q}_{\scriptscriptstyle H}=\pmatrix{0 & 0 & 0 & 0\cr \partial_q-\beta
\partial_qH & 0 & 0 & 0\cr 
\partial_p-\beta\partial_pH & 0 & 0
& 0\cr 0 & \partial_p-\beta\partial_pH & -\partial_q+\beta\partial_qH & 0}
\nonumber\\
\bigskip
&&\widehat{\overline{Q}}_{\scriptscriptstyle H}=\pmatrix{0 &
\partial_p+\beta\partial_pH & -\partial_q-\beta\partial_qH & 0\cr 0 & 0 & 0 & 
-\partial_q-\beta\partial_qH\cr 0 & 0 & 0 &
-\partial_p-\beta\partial_pH \cr 0 & 0 & 0 & 0}. \label{matrixn1}
\end{eqnarray}
Finally the matrix that represents the operator of evolution $\cal H$:
\begin{equation}
{\cal
H}=\widehat{L}+i\overline{c}_q\partial_p\partial_pHc^p+i\overline{c}_q\partial_p
\partial_qHc^q-i\overline{c}_p\partial_q\partial_pHc^p
-i\overline{c}_p\partial_q\partial_qHc^q,
\end{equation}
where $\widehat{L}=\lambda_a\omega^{ab}\partial_bH$ is the Liouville operator,
is given by:
\begin{equation}
\displaystyle 
\widehat{{\cal H}}=\pmatrix{\widehat{L} & 0 & 0 & 0\cr 0 & 
\widehat{L}-i\partial_q\partial_p H & i\partial_q\partial_qH & 0\cr
0 & -i\partial_p\partial_pH & \widehat{L}+i\partial_p\partial_qH & 0\cr 0 & 0 & 0 & 
\widehat{L}}.
\label{accatilde}
\end{equation}
From the expression of $\widehat{{\cal H}}$ above it is clear that the 0- and the 2-forms,
which in the 4-vector representation (\ref{genw2}), have respectively only the 
first and the last components different from zero, evolve only
with  the Liouvillian $\widehat{L}$. The 1-forms instead evolve with the central $2\times 2$ submatrix 
of Eq. (\ref{accatilde}) which contains all the possible second derivatives of the 
Hamiltonian $H(\varphi)$ and mixes the two central 
components
of the 4-vector $\psi$. For an example of how it is possible to reconstruct 
the evolution of the states at the matrix level the reader can consult Appendix {\bf A}.
It is easy to check that also at the matrix level the usual supersymmetry algebra,
$[\widehat{Q}_{\scriptscriptstyle H},\widehat{\overline{Q}}_{\scriptscriptstyle H}]_{\scriptscriptstyle +}=
2i\beta\widehat{{\cal H}}\;$, $[\widehat{Q}_{\scriptscriptstyle H},\widehat{\cal H}]_{\scriptscriptstyle -}=
[\widehat{\overline{Q}}_{\scriptscriptstyle H},\widehat{\cal H}]_{\scriptscriptstyle -}=0$, 
holds. So we have represented the superalgebra of the symmetry charges of the CPI
in terms of $4\times 4$ matrices of operators. Now 
irreducible representations of superalgebras are well-known in literature. 
In Appendix {\bf B} we shall use the results of Ref. \cite{Rittenberg} to
show how it is possible to build an irreducible representation for the superalgebra of the CPI
in terms of $4\times 4$ matrices whose entries are real numbers.

All the symmetries of the CPI turn the states into each other within the 
eigenspaces of the operator of evolution $\widehat{\cal H}$. For
example we can start considering an eigenstate of the Liouvillian $\widehat{L}$ 
with eigenvalue $l$: 
$\widehat{L}\psi_0^l=l\psi_0^l$. Since $\psi_0^l$ is a 0-form we can represent 
it by means of the following 4-vector:
\begin{equation}
\psi_0^l=\left( \begin{array}{c} \psi_0^l \\ 0 \\ 0 \\ 0\end{array}\right)
\end{equation}
which is an eigenstate for $\widehat{\cal H}$ with eigenvalue $l$: 
$\widehat{\cal H}\psi_0^l=l\psi_0^l$.
Since $\widehat{Q}$ commutes with $\widehat{\cal H}$ also $\widehat{Q}\psi_0^l$ 
is an eigenstate for $\widehat{\cal H}$ 
with the same eigenvalue $l$:
\begin{equation}
[\widehat{\cal H}, \widehat{Q}]=0\;\Longrightarrow\; \widehat{\cal H}
(\widehat{Q}\psi_0^l)=l(\widehat{Q}\psi_0^l).
\end{equation}
The explicit form of $\widehat{Q}\psi_0^l$ is given by the following 1-form:
\begin{equation}
\displaystyle 
\widehat{Q}\psi_0^l=\pmatrix{0 & 0 & 0 & 0\cr \partial_q & 0 & 0 & 0\cr \partial_p & 0 
& 0 & 0\cr 0 & \partial_p & -\partial_q & 0}
\left( \begin{array}{c} \psi_0^l\\ 0\\ 0\\ 0 \end{array} \right)=
\left( \begin{array}{c} 0\\ \partial_q\psi_0^l\\ \partial_p\psi_0^l\\ 0 
\end{array} \right). \label{BRS}
\end{equation}
In the same way $\widehat{K}\psi_0^l$ is an eigenstate for 
$\widehat{\cal H}$ with the same eigenvalue $l$. Its explicit form is given by:
\begin{equation}
\widehat{K}\psi_0^l=\pmatrix{0 & 0 & 0 & 0\cr 0 & 0 & 0 & 0\cr 0 & 0 & 0 & 0\cr 1 & 0 & 0 & 0}
 \left( \begin{array}{c}\psi_0^l\\ 0\\ 0\\ 0\end{array}\right)=
\left( \begin{array}{c}0\\ 0\\ 0\\ \psi_0^l\end{array}\right)
\end{equation}
and it is a 2-form.
So the symmetry charges move us within the eigenspace of $\widehat{\cal H}$, 
for example the one with eigenvalue 
$l$, passing from the 0-forms to the 1-forms
(by the $\widehat{Q}$ charge) and from the 0-forms to the 2-forms (by the 
$\widehat{K}$ charge). A similar role is played also by the supersymmetry charges. 
First of all, following Ref. \cite{Ioffe2}, we can rewrite the matrix realization of 
$Q_{\scriptscriptstyle H}, 
\overline{Q}_{\scriptscriptstyle H}$, that we derived in Eq. (\ref{matrixn1}), as:
\begin{equation}
\widehat{Q}_{\scriptscriptstyle H}=\pmatrix{0 & 0 & 0 & 0\cr 
Q_1^- & 0 & 0 & 0\cr Q_2^- & 0 & 0 & 0\cr 
0 & Q_2^- & -Q_1^- & 0}, \;\;\; \widehat{\overline{Q}}_{\scriptscriptstyle H}
=\pmatrix{0 & Q_1^+ & Q_2^+ & 0 \cr
0 & 0 & 0 & Q_2^+\cr 0 & 0 & 0 & -Q_1^+\cr 0 & 0 & 0 & 0} \label{susy}
\end{equation}
where 
\begin{eqnarray}
&&Q_1^-=\partial_q-\beta\partial_qH, \;\;\;\;\;\; Q_1^+=\partial_p+
\beta \partial_pH\nonumber\\
&&Q_2^-=\partial_p-\beta\partial_pH, \;\;\;\;\;\; Q_2^+=-\partial_q-
\beta\partial_qH.
\end{eqnarray}
Since $\widehat{Q}_{\scriptscriptstyle H}$ commutes with the Hamiltonian, $[\widehat{\cal H}, 
\widehat{Q}_{\scriptscriptstyle H}]=0$, we have that, if $\psi_{\scriptscriptstyle 0}^l$
is an eigenstate for the Liouvillian, then 
\begin{equation}
\widehat{Q}_{\scriptscriptstyle H} \left( \begin{array}{c}\psi_0^l\\ 0\\ 0\\ 0\end{array}\right)
=\left( \begin{array}{c}0\\ Q_1^-\psi_{\scriptscriptstyle 0}^l\\ Q_2^-\psi_{\scriptscriptstyle 0}^l\\ 0\end{array}\right)
\label{3-19}
\end{equation} 
is also an eigenstate for $\widehat{\cal H}$ with the same eigenvalue $l$. Not only, but if we rewrite 
the operator of evolution (\ref{accatilde}) as:
\begin{equation}
\widehat{\cal H}=\pmatrix{\widehat{L} & 0 & 0\cr 0 & 
\widehat{\cal H}_{kj}^{\scriptscriptstyle (1)} & 0\cr 0 & 0 & \widehat{L}}
\end{equation}
where
\begin{equation}
\widehat{\cal H}^{\scriptscriptstyle (1)}=\pmatrix{\widehat{L}-i\partial_q\partial_pH & 
i\partial_q\partial_qH\cr -i\partial_p\partial_p H & \widehat{L}
+i\partial_p\partial_qH}
\end{equation}
then the following relation holds:
\begin{equation}
\widehat{\cal H}^{\scriptscriptstyle (1)}\left(\begin{array}{c}Q_1^-
\psi_{\scriptscriptstyle 0}^l\\ Q_2^-\psi_{\scriptscriptstyle 0}^l
\end{array}\right)=
l\left(\begin{array}{c} Q_1^-\psi_{\scriptscriptstyle 0}^l \\ Q_2^-
\psi_{\scriptscriptstyle 0}^l\end{array}
\right) \label{3-22}
\end{equation}
i.e. the 2-vector $Q_j^-\psi_{\scriptscriptstyle 0}^l$ is 
eigenstate for $\widehat{\cal H}^{\scriptscriptstyle (1)}$ with eigenvalue
$l$:
\begin{equation}
\widehat{\cal H}^{\scriptscriptstyle (1)}_{kj}(Q_j^-\psi_{\scriptscriptstyle 0}^l)=l(Q_k^-\psi_{\scriptscriptstyle 0}^l)
\label{3-23}.
\end{equation}
Vice versa if a state $\psi_{k}^{\scriptscriptstyle (1)}$ is an eigenstate of the operator 
$\widehat{\cal H}^{\scriptscriptstyle (1)}$ with eigenvalue $l$, we have then that
the associated 4-vector $\displaystyle \left(\begin{array}{c} 0 \\ \psi_{k}^{\scriptscriptstyle (1)} 
\\ 0\end{array}\right)$ is
an eigenstate for $\widehat{\cal H}$ with the same eigenvalue. 
Next let us notice how $\widehat{\overline{Q}}_{\scriptscriptstyle H}$ 
acts on states of the form $\left(\begin{array}{c}
0 \\ \psi_{1}^{\scriptscriptstyle (1)}\\ \psi_{2}^{\scriptscriptstyle (1)} \\ 0\end{array}\right)$:
\begin{equation}
\widehat{\overline{Q}}_{\scriptscriptstyle H}\left(\begin{array}{c} 0 \\
\psi_{k}^{\scriptscriptstyle (1)} \\ 0\end{array}\right)=
\pmatrix{0 & Q_1^+ & Q_2^+ & 0\cr
0 & 0 & 0 & Q_2^+ \cr 0 & 0 & 0 & -Q_1^+\cr 0 & 0 & 0 & 0}\cdot\left(\begin{array}{c}
0 \\ \psi_{1}^{\scriptscriptstyle (1)}\\ \psi_{2}^{\scriptscriptstyle (1)} \\ 0\end{array}\right)=\left(\begin{array}{c} 
Q_k^+\psi_{k}^{\scriptscriptstyle(1)}\\ 0 \\ 0\\ 0
\end{array}\right). \label{3.24}
\end{equation}
As $\widehat{\overline{Q}}_{\scriptscriptstyle H}$ commutes with $\widehat{\cal H}$, we can conclude
from (\ref{3.24}) that the two states $\displaystyle \left(\begin{array}{c} 0 \\ \psi_{k}^{\scriptscriptstyle (1)} 
\\ 0\end{array}\right)$ and $\left(\begin{array}{c} 
Q_k^+\psi_{k}^{\scriptscriptstyle(1)}\\ 0 \\ 0\\ 0
\end{array}\right)$ are degenerate. From the form of these two states we can also phrase this degeneracy
by saying that, if $\psi_k^{\scriptscriptstyle (1)}$ is an eigenstate of $\widehat{\cal H}^{\scriptscriptstyle (1)}$,
then the operators $Q_k^+$ map the eigenstates of 
$\widehat{\cal H}^{\scriptscriptstyle (1)}$ into eigenstates of the Liouvillian
$\widehat{L}$, according to the following relation:
\begin{equation}
\widehat{L}(Q_k^+\psi_{k}^{\scriptscriptstyle (1)})=
l(Q_k^+\psi_{k}^{\scriptscriptstyle (1)})
\end{equation}
where summation over $k$ is understood.
A more complete and refined analysis can be performed on the basis of what has been done for supersymmetric
quantum mechanics in Ref. \cite{Ioffe2}. The final result is that the 
two quite different operators like $\widehat{L}$ and $\widehat{\cal H}^{\scriptscriptstyle (1)}$
have equivalent spectra. The only difference might be in the handling of the zero eigenvalue.
We should notice that in our case one of the two operators, the Liouvillian $\widehat{L}$, has a deep 
physical meaning and its spectrum gives us information on important properties like the ergodicity 
\cite{Arnold}, the mixing of the system, etc. So the fact that its spectrum
is equivalent to the one of $\widehat{\cal H}^{\scriptscriptstyle (1)}$ may help in discovering further 
things on dynamical systems. 

Up to now we have specified only which is the space of the wave functions whose evolution
is given by the CPI, i.e. the space of 4-vectors $\psi$ of Eq. (\ref{genw2}). 
To build a true Hilbert space we have to introduce also a
suitable scalar product between two different wave functions $\psi$ and $\Phi$. Following Ref. \cite{Salomonson}
one of the most natural choices is:
\begin{equation}
\displaystyle \langle \psi|\Phi\rangle =\int d\varphi[\psi_{\scriptscriptstyle 0}^*\Phi_{\scriptscriptstyle 0}
+\psi_q^*\Phi_q+\psi_p^*\Phi_p+\psi^*_{\scriptscriptstyle 2}\Phi_{\scriptscriptstyle 2}]. \label{scp}
\end{equation} 
With this scalar product all the states have positive definite norms and the only state with zero norm is the null state.
It is possible to rewrite the scalar product (\ref{scp}) as:
\begin{equation}
\displaystyle \langle \psi|\Phi\rangle =\int d\varphi (\psi_{\scriptscriptstyle 0}^* \,\psi_q^* \,\psi_p^*\,
\psi_{\scriptscriptstyle 2}^*)\pmatrix{1 & 0 & 0 & 0\cr 0 & 1 & 0 & 0\cr 0 & 0 & 1 & 0\cr 0 & 0 & 0 & 1}
\left( \begin{array}{c} \Phi_{\scriptscriptstyle 0}\\ \Phi_q\\ \Phi_p\\ \Phi_{\scriptscriptstyle 2}
\end{array} \right) \label{314}
\end{equation}
or, in a more compact way:
\begin{equation}
\displaystyle \langle \psi|\Phi\rangle =\int d\varphi \,(\psi^{\scriptscriptstyle T})^*
\cdot\Phi.
\end{equation}
With this inner product it is easy to prove that:
\begin{equation}
\langle \psi|c^a\Phi\rangle =\langle\overline{c}_a\psi|\Phi\rangle,\;\;\;\;\; \langle \psi|\overline{c}_a\Phi\rangle
=\langle c^a\psi|\Phi\rangle \label{3-31}
\end{equation}
i.e. the $c$ and $\overline{c}$ operators are one the hermitian conjugate of the other:
$\overline{c}=c^{\dagger}=(c^{\scriptscriptstyle T})^*$.
With this rules the two number operators $N_q=c^q\overline{c}_q$ and $N_p=c^p\overline{c}_p$ are hermitian and commute
with each other. Since $N_q^2=N_q$ and $N_p^2=N_p$ the only possible eigenvalues are 0 and 1 as it is 
particularly clear using their matrix representation derived from the matrix representation 
of $c^q$ and $c^p$:
\begin{eqnarray}
&&\widehat{N}_q=\widehat{c}^q\widehat{\overline{c}}_q=
\pmatrix{0 & 0 & 0 & 0\cr 1 & 0 & 0 & 0\cr 0 & 0 & 0 & 0\cr 0 & 0 & 1 &
0}\cdot
\pmatrix{0 & 1 & 0 & 0\cr 0 & 0 & 0 & 0\cr 0 & 0 & 0 & 1\cr 0 & 0 & 0 & 0}=
\pmatrix{0 & 0 & 0 & 0\cr 0 & 1 & 0 & 0\cr 0 & 0 & 0 & 0\cr 0 & 0 & 0 & 1}\\
&&\widehat{N}_p=\widehat{c}^p\widehat{\overline{c}}_p=
\pmatrix{0 & 0 & 0 & 0\cr 0 & 0 & 0 & 0\cr 1 & 0 & 0 & 0\cr 0 & -1 & 0 & 0}\cdot
\pmatrix{0 & 0 & 1 & 0\cr 0 & 0 & 0 & -1\cr 0 & 0 & 0 & 0\cr 0 & 0 & 0 & 0}=
\pmatrix{0 & 0 & 0 & 0\cr 0 & 0 & 0 & 0\cr 0 & 0 & 1 & 0\cr 0 & 0 & 0 & 1}.\nonumber
\end{eqnarray}
$\widehat{N}_q$ and $\widehat{N}_p$ are a complete set of commuting and hermitian operators 
for what concerns the Grassmannian part of
the theory. This means that the knowledge of the simultaneous eigenvalues of $\widehat{N}_q$ and 
$\widehat{N}_p$ allows us to specify in a
unique way which of the 4 basis vectors we have to consider. 
The correspondence between the eigenvalues $(n_q,n_p)$ of
$(\widehat{N}_q,\widehat{N}_p)$ and the basis state vectors of (\ref{genw2}) is given by the following table:
\begin{eqnarray}
&& (0,0)\;\Longleftrightarrow\;\left( \begin{array}{c} 1\\ 0\\
0\\ 0\end{array}\right),\;\;\;\;\;\;\;\;\;\; (1,0)\;\Longleftrightarrow\;\left(
\begin{array}{c} 0\\ 1 \\ 0\\ 0\end{array}\right)\nonumber\\
&& (0,1)\;\Longleftrightarrow\; \left( \begin{array}{c} 0\\ 0\\ 1\\
0\end{array}\right),\;\;\;\;\;\;\;\;\;\; (1,1)\;\Longleftrightarrow\;
\left( \begin{array}{c} 0\\ 0\\ 0\\ 1\end{array}\right).
\end{eqnarray}
Therefore every wave function $\psi$ of the
generalized Hilbert space can be expanded on the basis of the common eigenstates of 
$\widehat{N}_q$ and $\widehat{N}_p$ and it is possible to construct a resolution of the identity involving only 
these eigenstates. See also Ref. \cite{dgm} where other types of scalar products have been analyzed.

\section{Cartan Calculus in the $n=1$ Case}

It is well-known from the original papers \cite{Gozzi} that a lot of operations
of the Cartan calculus can be performed via the symmetry charges of the CPI. In this paper
we have seen how, in the case $n=1$, all these symmetry charges can be represented
via $4\times 4$ matrices. Therefore we expect that also the operations of
differential geometry can be performed in terms of $4\times 4$ matrices. We 
start remembering that if $n=1$ the phase space is labeled by two variables $q$ and $p$,
a basis for the cotangent bundle is given by $dq\equiv c^q$ and $dp\equiv c^p$ and, as we have seen 
in Eq. (\ref{genw2}), the most general non-homogeneous differential form 
$\psi=\psi_{\scriptscriptstyle 0}+\psi_qc^q+\psi_pc^p+\psi_{\scriptscriptstyle 2}c^pc^q$ 
can be represented by the 4-vector $\displaystyle \psi=\left( \begin{array}{c} 
\psi_{\scriptscriptstyle 0}\\ \psi_q\\ \psi_p\\ \psi_{\scriptscriptstyle 2} \end{array} \right)$. 

\medskip

\noindent $\bullet$ {\bf Exterior Derivative}. 
With the identifications $dq\equiv c^q$ and $dp\equiv c^p$ we have that, for $n=1$,
the action of the exterior derivative {\bf d} on a 0-form is given by
${\bf d}\psi_{\scriptscriptstyle 0}=\partial_q\psi_{\scriptscriptstyle 0}c^q
+\partial_p\psi_{\scriptscriptstyle 0}c^p$; on a 1-form is given by 
${\bf d}(\psi_qc^q+\psi_pc^p)=(\partial_p\psi_q-\partial_q\psi_p)c^pc^q$
and finally on a 2-form is simply ${\bf d}(\psi_{\scriptscriptstyle 2}
c^pc^q)=0$. The symmetry charge
$\widehat{Q}$ can be interpreted as the exterior derivative also at the matrix level. 
In fact when we apply the matrix $\widehat{Q}$ over the 4-vector
$\psi$ we produce a new 4-vector whose components are just the 
4 components of the differential form ${\bf d}\psi$ obtained acting with the exterior derivative
${\bf d}$ over $\psi$.
\begin{equation}
\widehat{Q}\psi=\pmatrix{0 & 0 & 0 & 0\cr \partial_q & 0 & 0 & 0\cr \partial_p & 
0 & 0 & 0\cr 0 & \partial_p & -\partial_q & 0}
\left( \begin{array}{c} \psi_{\scriptscriptstyle 0}\\ \psi_q\\ \psi_p\\ \psi_{\scriptscriptstyle 2} 
\end{array} \right)=
\left( \begin{array}{c} 0 \\ \partial_q \psi_{\scriptscriptstyle 0} \\ \partial_p \psi_
{\scriptscriptstyle 0} \\ \partial_p
\psi_q-\partial_q\psi_p \end{array} \right)
\equiv {\bf d}\psi. \label{extder}
\end{equation}

\medskip

\noindent $\bullet$ {\bf Form Number}. The symmetry charge $\widehat{Q}_f$ 
is in relation with the form number of $\psi$. In fact:
\begin{equation}
\widehat{Q}_f\psi^{\scriptscriptstyle (p)}=\pmatrix{0 & 0 & 0 & 0\cr 0 & 1 & 0 & 0\cr 0 & 0 & 1 & 0\cr 
0 & 0 & 0 & 2}\left( \begin{array}{c}
\psi^{\scriptscriptstyle (0)}_{\scriptscriptstyle 0} \\ \psi_q^{\scriptscriptstyle (1)} \\ 
\psi_p^{\scriptscriptstyle (1)} \\ \psi_{\scriptscriptstyle 2}^{\scriptscriptstyle (2)}\end{array}\right)
=\left( \begin{array}{c}
0\cdot \psi^{\scriptscriptstyle (0)}_{\scriptscriptstyle 0} \\ 1\cdot\psi_q^{\scriptscriptstyle (1)} \\ 
1\cdot\psi_p^{\scriptscriptstyle (1)} \\ 2\cdot\psi_{\scriptscriptstyle 2}^{\scriptscriptstyle (2)}\end{array}\right)
=p\psi^{\scriptscriptstyle (p)}
\end{equation}
where the previous relation means that all the homogeneous forms are eigenstates
for $\widehat{Q}_f$. In particular the 0-forms $\psi^{\scriptscriptstyle (0)}$
are eigenstates of $\widehat{Q}_f$ 
with eigenvalue 0, the 1-forms $\psi^{\scriptscriptstyle (1)}$ are eigenstates
with eigenvalue 1 and the 2-forms $\psi^{\scriptscriptstyle (2)}$ are eigenstates with eigenvalue 2. 

\medskip

\noindent $\bullet$ {\bf Interior Contraction}.
Which is the result of the interior contraction of a generic form with the 
vector field $V=V^a\overline{c}_a
=V^q\overline{c}_q+V^p\overline{c}_p$? Every 0-form has 
interior contraction 0 with $V$: 
$\iota_{\scriptscriptstyle V}\psi^{\scriptscriptstyle (0)}=0$. Every 1-form of the type 
$\psi^{\scriptscriptstyle (1)}=\psi_q^{\scriptscriptstyle (1)}c^q+\psi_p^{
\scriptscriptstyle (1)}c^p$ 
has the following interior 
contraction: $\iota_{\scriptscriptstyle V}\psi^{\scriptscriptstyle (1)}=
V^q\psi^{\scriptscriptstyle (1)}_q+V^p\psi^{\scriptscriptstyle (1)}_p$.
Finally the interior contraction with $V$ of the 2-form 
$\psi^{\scriptscriptstyle (2)}=\psi^{\scriptscriptstyle (2)}_{\scriptscriptstyle 2}c^pc^q$ is given by:
$\iota_{\scriptscriptstyle V}\psi^{\scriptscriptstyle (2)}=V^p\psi^{\scriptscriptstyle (2)}
_{\scriptscriptstyle 2}c^q-V^q
\psi^{\scriptscriptstyle (2)}_{\scriptscriptstyle 2}c^p$. 
So we can say that the interior contraction of a form with the vector field $V$
maps the 4-vector $\displaystyle \left( \begin{array}{c} \psi^{\scriptscriptstyle (0)}_{\scriptscriptstyle 0}\\ 
\psi_q^{\scriptscriptstyle (1)}\\ \psi_p^{\scriptscriptstyle (1)}\\ 
\psi_{\scriptscriptstyle 2}^{\scriptscriptstyle (2)} \end{array}\right)$ 
into the 4-vector $\displaystyle \left( \begin{array}{c} 
V^q\psi_q^{\scriptscriptstyle (1)}+V^p\psi_p^{\scriptscriptstyle (1)}\\ 
V^p\psi_{\scriptscriptstyle 2}^{\scriptscriptstyle (2)} \\ -V^q\psi_{\scriptscriptstyle 2}^{\scriptscriptstyle (2)}\\ 0
\end{array}\right)$.  It is easy to see that the matrix that realizes the previous mapping 
is given by:
\begin{equation}
\displaystyle 
\iota_{\scriptscriptstyle V}=\pmatrix{0 & V^q & V^p & 0\cr 0 & 0 & 0 &
V^p\cr 0 & 0 & 0 & -V^q\cr 0 & 0 & 0 & 0}. \label{intcon}
\end{equation}
The matrix (\ref{intcon}) is just equal to $\widehat{V}=V^q\widehat{\overline{c}}_q+V^p
\widehat{\overline{c}}_p$ where $\widehat{\overline{c}}_q$ and 
$\widehat{\overline{c}}_p$
are the matrix representations of the fermionic operators, see formula (\ref{operator3}). 

\medskip

\noindent $\bullet$ {\bf Lie Derivative along the Hamiltonian Flow}. 
If we take, as a particular case of the previous analysis,
a Hamiltonian vector field $h^a=\omega^{ab}\partial_bH$, we have that 
$V^q=\partial_pH$, $V^p=-\partial_qH$
and the interior contraction (\ref{intcon}) becomes:
\begin{equation}
\iota_{h}=\pmatrix{0 & \partial_pH & -\partial_qH & 0\cr 0 & 0 & 0 & 
-\partial_qH\cr 0 & 0 & 0 & -\partial_pH\cr
0 & 0 & 0 & 0}. \label{iota}
\end{equation}
From the matrix representation of the interior contraction (\ref{iota}) 
and of the exterior derivative (\ref{extder}) we can easily derive the matrix representation
for the Lie derivative associated with the Hamiltonian vector field $h$:
\begin{equation}
\displaystyle 
{\cal L}_h={\bf d}\iota_h+\iota_h{\bf d}=\pmatrix{i\widehat{L} & 0 & 0 & 0\cr 0 & i\widehat{L}+
\partial_q\partial_pH & -\partial_q\partial_qH & 0\cr
0 & \partial_p\partial_pH & i\widehat{L}-\partial_p\partial_qH & 0\cr 0 & 0 & 
0 & i\widehat{L}}. \label{lieder}
\end{equation}
By comparing (\ref{accatilde}) with (\ref{lieder}) we have that 
also at the matrix level
$\widehat{\cal H}=-i{\cal L}_h$. So, when we apply the operator of 
evolution 
$\widehat{\cal H}$ on a generic 
form $\psi$ we obtain a form whose components are given, modulus a factor 
$-i$, by the Lie 
derivative of the Hamiltonian flow: $\widehat{\cal H}\psi=-i{\cal L}_h\psi$.

\medskip

\noindent $\bullet$ {\bf Hodge Star}. The Hodge $*$ transformation 
is defined as \cite{Eguchi}:
\begin{equation}
\displaystyle 
*(dx^{i_1}\wedge dx^{i_2}\wedge \ldots \wedge dx^{i_p})=\frac{1}{(n-p)!}
\epsilon_{i_1i_2\ldots i_pi_{p+1}\ldots i_N}
dx^{i_{p+1}}\wedge dx^{i_{p+2}}\wedge\cdots \wedge dx^{i_N}.
\end{equation}
In the language of the CPI we have that in the case of one degree of freedom ($N=2$): 
\begin{eqnarray}
&&*(1)=\epsilon_{qp} dq\wedge dp=dq\wedge dp\;\;\Longrightarrow\;\; *(1)=c^qc^p\nonumber\\
&&*(dx^i)=\epsilon_{ij}dx^j\;\;\Longrightarrow\;\; *(c^q)=c^p,\;\,*(c^p)=-c^q\nonumber\\
&&*(dp\wedge dq)=\epsilon_{pq}=-1\;\;\Longrightarrow\;\; *(c^pc^q)=-1.
\end{eqnarray}
At the matrix level, with the convention (\ref{genw2}), the action of $*$ is given by:
\begin{eqnarray}
&&*\left( \begin{array}{c} 1\\0\\0\\0\end{array} \right)=
\left( \begin{array}{c} 0\\0\\0\\-1\end{array} \right),\qquad\quad
*\left( \begin{array}{c} 0\\1\\0\\0\end{array} \right)=\left( \begin{array}{c} 
0\\0\\1\\0\end{array} \right),\nonumber\\
&&*\left( \begin{array}{c} 0\\0\\1\\0\end{array} \right)=\left( \begin{array}{c} 
0\\-1\\0\\0\end{array} \right),\qquad\quad
*\left( \begin{array}{c} 0\\0\\0\\1\end{array} \right)=\left( 
\begin{array}{c} -1\\0\\0\\0\end{array} \right).
\end{eqnarray}
Therefore the matrix representation for the Hodge $*$ transformation is:
\begin{equation}
\displaystyle
*=\pmatrix{0 & 0 & 0 & -1\cr 0 & 0 & -1 & 0\cr 0 & 1 & 0 & 0\cr -1 & 0 & 0 & 0}. \label{hodge}
\end{equation}

\medskip

\noindent $\bullet$ {\bf The adjoint of {\bf d}}. The $*$ transformation can be used
to define the following ``scalar product" between forms:
\begin{equation}
\displaystyle
(\alpha_p,\beta_p)=\int_{\cal M} \alpha_p\wedge *\beta_p. \label{innpro}
\end{equation}
Using it we can then define the adjoint $\delta$ of the exterior derivative {\bf d} 
as:
\begin{equation}
(\alpha_p,{\bf d}\beta_{p-1})\equiv(\delta\alpha_p,\beta_{p-1}). \label{4-11}
\end{equation}
In particular it is possible to prove \cite{Eguchi} that, in the case of a manifold with 
even dimension,
the relation between $\delta$ and ${\bf d}$ is $\delta=-*{\bf d}*$.
From the matrix representation of ${\bf d}$, Eq. (\ref{extder}), and of $*$, Eq. (\ref{hodge}), 
we obtain the following matrix representation for $\delta$:
\begin{equation}
\delta=\pmatrix{0 & -\partial_q & -\partial_p & 0\cr 0 & 0 & 0 & 
-\partial_p\cr 0 & 0 & 0 &\partial_q\cr 
0 & 0 & 0 & 0}.
\end{equation}
$\delta$, as well as ${\bf d}$, is a nilpotent matrix and
it lowers the degree of the forms by one. So somehow it acts like 
the symmetry charge $\widehat{\overline{Q}}$,
but, nevertheless, it does not coincide with it as we will prove now. 
In the language of the CPI we have that $\partial_a=i\lambda_a$ and, 
from the explicit form of the matrices $\widehat{\overline{c}}$, Eq. (\ref{operator3}),
we easily obtain that:
\begin{equation}
\delta=\pmatrix{0 & -i\lambda_q & -i\lambda_p & 0\cr 0 & 0 & 0 & -i\lambda_p\cr 
0 & 0 & 0 & i\lambda_q\cr 0 & 0 & 0 & 0}\equiv -i\lambda_q\widehat{\overline{c}}_q-i
\lambda_p\widehat{\overline{c}}_p. \label{delta}
\end{equation}
Note that the RHS of (\ref{delta}) is not $\overline{Q}$ which is $\overline{Q}=-i\lambda_q\overline{c}_p
+i\lambda_p\overline{c}_q$. 
Before concluding this section let us notice that the ``inner product" (\ref{innpro}) used in 
differential geometry
coincides with the positive definite inner product defined in Eq. (\ref{scp}). 
Moreover from the associated hermiticity relations (\ref{3-31}) among the Grassmannian operators
$c^{q^{\dagger}}=\overline{c}_q,\;c^{p^{\dagger}}=\overline{c}_p$ we have that
\begin{equation} 
\delta=(ic^q\lambda_q+ic^p\lambda_p)^{\dagger}\;\Longrightarrow\;\delta={\bf d}^{\dagger}
\end{equation}
which is just equivalent to relation (\ref{4-11}).

\medskip

\noindent $\bullet$ {\bf The Laplacian}. The Laplacian is defined 
starting from ${\bf d}$ and $\delta$
as:
\begin{equation}
\Delta=({\bf d}+\delta)^2={\bf d}\delta+\delta{\bf d}.
\end{equation} 
From (\ref{extder}) and (\ref{delta}) we have that the matrix representation of 
the Laplacian is given by ${\bf d}\delta+\delta {\bf d}=(-\partial_q^2-
\partial_p^2){\bf 1}_{\scriptscriptstyle 4\times 4}$ or, in the language of the CPI \cite{Gozzi}:
\begin{equation}
\Delta=(\lambda_q^2+\lambda_p^2){\bf 1}_{\scriptscriptstyle 4\times 4}
\end{equation}
from which it is particularly clear that $\Delta$ is a positive definite operator.

\section{Grassmann Algebras and Pauli Matrices}

In order to generalize all the results of the previous sections to the case 
of a system with an arbitrary great number of degrees of freedom, it is particularly useful to find a
representation of the Grassmannian operators in terms of tensor products of Pauli 
matrices. For the
kind of analysis which follows I am greatly indebted to M.V. Ioffe \cite{Ioffe}.
In the case of $n=1$ we had identified $\displaystyle \widehat{c}^p=\frac{\sigma^
{\scriptscriptstyle (-)}}{2}\otimes {\bf 1}$ and $\displaystyle \widehat{c}^q=\sigma_z\otimes
\frac{\sigma^{\scriptscriptstyle(-)}}{2}$, in the general case we will re-arrange
the variables $\varphi^a$ as: 
$\varphi^1=p_1,\,\varphi^2=q_1,\,\varphi^3=p_2,\,\varphi^4=q_2$ and so on. Using this convention for the indices
$1,2,\ldots, k$ on $\varphi^k$ and $c^k$, the correspondence between the Grassmann operators and the Pauli
matrices is:
\begin{equation}
\left\{
	\begin{array}{l}
	\displaystyle
	\label{general}
	\widehat{c}^k=(\sigma_z)^{\otimes k-1}\otimes \frac{\sigma^{
        \scriptscriptstyle (-)}}{2}
        \otimes ({\bf 1})^{\otimes 2n-k}, \;\;\;\;\; k=1,\cdots, 2n \smallskip\\
	\displaystyle \widehat{\overline{c}}_j=(\sigma_z)^{\otimes j-1}\otimes\frac{\sigma^{
        \scriptscriptstyle (+)}}{2}\otimes({\bf 1})^{\otimes 2n-j}, \;\;\;\;\; j=1,\cdots,2n\\
	\end{array} \label{5-1}
	\right.
\end{equation}
where
\begin{equation}
(\sigma_z)^{\otimes k-1}=\underbrace{\sigma_z\otimes \sigma_z \otimes
\cdots \otimes \sigma_z}_{k-1 \;times}
\end{equation}
and similarly for the other Pauli or identity matrices.

The next thing to do is to check that the $\widehat{c}^k,\widehat{\overline{c}}_j$ 
built in (\ref{5-1}) satisfy the usual algebra of the Grassmannian variables
\begin{equation}
[\widehat{c}^a,\widehat{\overline{c}}_b]_+=\delta^a_b, \;\;\;\;\;\; 
[\widehat{c}^a,\widehat{c}^b]_+=[\widehat{\overline{c}}_a,\widehat{\overline{c}}_b]_+=0,\;\;\;\; a,b=1,\dots, 2n
\label{5-3}.
\end{equation}
The calculation is quite long and it is reported in Appendix {\bf C}. Its basic ingredients are the usual 
properties of the Pauli matrices $\sigma_x^2=\sigma_y^2=\sigma_z^2={\bf 1}_{\scriptscriptstyle 2\times 2}$,
$[\sigma_i,\sigma_j]_+=0$, $\displaystyle \biggl[\frac{\sigma^{\scriptscriptstyle (-)}}{2},
\frac{\sigma^{\scriptscriptstyle (+)}}{2}\biggr]_+={\bf
1}_{\scriptscriptstyle 2\times 2}$, $[\sigma^{\scriptscriptstyle (-)},
\sigma^{\scriptscriptstyle (-)}]_+=[\sigma^{\scriptscriptstyle (+)},\sigma^{\scriptscriptstyle (+)}]_+=0$ 
and the following other relation involving tensor products of Pauli matrices
\begin{equation}
[(a\otimes b),(c\otimes d)]=(a\cdot c)\otimes (b\cdot d)-(c\cdot a)\otimes (d\cdot b).
\end{equation}
So, thanks to the construction (\ref{general}), it is possible to give an expression for the Grassmann
operators of the CPI in terms of tensor
products of suitable Pauli or identity matrices. The Grassmann algebra of $\widehat{c},\widehat{\overline{c}}$ 
becomes then 
a direct consequence of the algebra of 
Pauli matrices. The construction (\ref{general}) will allow us to generalize all 
the results obtained in the case $n=1$
to the case of an arbitrary great number of degrees of freedom, without 
losing a certain compactness in the
appearance of the formulas. 

As we want to represent the Grassmann operators of the theory as tensor 
products of Pauli matrices we have to 
represent also the states of the Hilbert space underlying the CPI
as tensor products of 
2-dimensional vectors in a consistent way. 
For example in the
$n=1$ case, where we represented
$\displaystyle \widehat{c}^p=\frac{\sigma^{\scriptscriptstyle (-)}}{2}\otimes 
{\bf 1}$ and $\displaystyle \widehat{c}^q=\sigma_z\otimes 
\frac{\sigma^{\scriptscriptstyle (-)}}{2}$,
the correspondent Hilbert space can be constructed from all the possible tensor products  
of the 2-dimensional vectors on which the Pauli matrices act. These two basic vectors
could be $\displaystyle \left( \begin{array}{c} 1\\0 
\end{array} \right)$ and $\displaystyle \left( \begin{array}{c} 0\\1 \end{array} \right)$.
Their tensor products are:
\begin{eqnarray}
&&\left( \begin{array}{c} 1\\ 0 \end{array}\right)\otimes \left( 
\begin{array}{c} 1 \\ 0 \end{array}\right)=
\left( \begin{array}{c} 1 \\ 0 \\ 0 \\ 0 \end{array}\right)\Leftrightarrow 1,\qquad \quad
\left( \begin{array}{c} 1\\ 0 \end{array}\right)\otimes \left( 
\begin{array}{c} 0 \\ 1 \end{array}\right)=
\left( \begin{array}{c} 0 \\ 1 \\ 0 \\ 0 \end{array}\right)\Leftrightarrow c^q,\nonumber\\
&&\left( \begin{array}{c} 0\\ 1 \end{array}\right)\otimes \left( 
\begin{array}{c} 1 \\ 0 \end{array}\right)=
\left( \begin{array}{c} 0 \\ 0 \\ 1 \\ 0 \end{array}\right)\Leftrightarrow c^p,\qquad\;\;
\left( \begin{array}{c} 0\\ 1 \end{array}\right)\otimes \left( 
\begin{array}{c} 0 \\ 1\end{array} \right)=
\left( \begin{array}{c} 0 \\ 0 \\ 0 \\ 1 \end{array}\right)\Leftrightarrow c^pc^q.
\label{basis}
\end{eqnarray}
These four states and the identifications we have indicated on their RHS 
are obviously consistent with the expression 
(\ref{genw2}) since it is from there 
that we began our analysis. What we mean is that from (\ref{basis}) we obtain 
that the generic form
\begin{equation}
\psi(\varphi,c)=\psi_{\scriptscriptstyle 0}(\varphi)\cdot 1
+\psi_q(\varphi)\cdot c^q+\psi_p(\varphi)\cdot c^p+
\psi_{\scriptscriptstyle 2}(\varphi)\cdot c^pc^q \label{5-7}
\end{equation}
can be identified with the 4-vector:
\begin{equation}
\psi(\varphi,c)=\left(\begin{array}{c} \psi_{\scriptscriptstyle 0}(\varphi) \\ \psi_q(\varphi) 
\\ \psi_p(\varphi) \\ \psi_{\scriptscriptstyle 2}(\varphi)
\end{array} \right) 
\end{equation}
that is just the RHS of Eq. (\ref{genw2}).
At this level an important thing to underline is that, once we have fixed the matrix 
representation of the Grassmannian operators, the representation of the 
states must be derived by consistency.
Therefore the choice (\ref{general}) implies that we have 
to order the components of the generic form $\psi$ 
in a very peculiar manner. For example suppose we take 
$n=2$ (that means $\varphi^a=(p^{\scriptscriptstyle 1},q^{\scriptscriptstyle 1},p^{\scriptscriptstyle 2},
q^{\scriptscriptstyle 2})$, $c^a=(c^{\scriptscriptstyle p_1},c^{\scriptscriptstyle q_1},c^{\scriptscriptstyle p_2},
c^{\scriptscriptstyle q_2})$, $\overline{c}_a=(\overline{c}_{\scriptscriptstyle p_1},
\overline{c}_{\scriptscriptstyle q_1},\overline{c}_{\scriptscriptstyle p_2},
\overline{c}_{\scriptscriptstyle q_2})$). The basis of the 0-form, which in (\ref{5-7}) is indicated with 1,
is given by the following vector with 16 components:
\begin{equation}
1=\left(\begin{array}{c} 1\\0\end{array}\right)\otimes 
\left(\begin{array}{c} 1\\0 \end{array} \right)\otimes 
\left(\begin{array}{c} 1\\0 \end{array} \right) \otimes 
\left(\begin{array}{c} 1\\0 \end{array} \right)=\delta_{i,1}.
\end{equation}
For reasons of space we have not written down explicitly the 16-components vector. We have indicated it
with $\delta_{i,1}$ which means a vector with an element $1$ in the first position and all the other 15
elements equal to $0$. 
For the 1-forms we have:
\begin{eqnarray}
&& c^{q_2}\equiv dq_2\Leftrightarrow\left(\begin{array}{c} 1\\0\end{array}\right)\otimes 
\left(\begin{array}{c} 1\\0\end{array}\right)\otimes
\left(\begin{array}{c} 1\\0\end{array}\right)\otimes\left(\begin{array}{c} 
0\\1\end{array}\right)
=\delta_{i,2},\nonumber\\
&& c^{p_2}\equiv dp_2\Leftrightarrow\left(\begin{array}{c} 1\\0\end{array}\right)\otimes 
\left(\begin{array}{c} 1\\0\end{array}\right)\otimes
\left(\begin{array}{c} 0\\1\end{array}\right)\otimes\left(\begin{array}{c} 
1\\0\end{array}\right)
=\delta_{i,3},\nonumber\\
&& c^{q_1}\equiv dq_1\Leftrightarrow\left(\begin{array}{c} 1\\0\end{array}\right)\otimes 
\left(\begin{array}{c} 0\\1\end{array}\right)\otimes
\left(\begin{array}{c} 1\\0\end{array}\right)\otimes\left(\begin{array}{c} 
1\\0\end{array}\right)
=\delta_{i,5},\nonumber\\
&& c^{p_1}\equiv dp_1\Leftrightarrow\left(\begin{array}{c} 0\\1\end{array}\right)\otimes 
\left(\begin{array}{c} 1\\0\end{array}\right)\otimes
\left(\begin{array}{c} 1\\0\end{array}\right)\otimes\left(\begin{array}{c} 
1\\0\end{array}\right)
=\delta_{i,9}. \label{5-10}
\end{eqnarray}
For the details of their construction see Appendix {\bf D}.
If we perform explicitly the previous tensor products we obtain four 
16-dimensional vectors with all the elements 
equal to 0 except for one element equal to 1 placed respectively in the second, 
the third, the fifth and the ninth
position. Here we note one defect of the representation we introduced: 
the components of the 1-form are scattered 
inside the 16-dimensional vector and they do not form a unique block of 
adjacent components, from the second to the 
fifth one, like it happens in the $n=1$ case.

Having represented the states as tensor products, it is quite evident 
the reason why  all the Grassmann
algebra can be reconstructed starting from $\displaystyle 
\frac{\sigma^{\scriptscriptstyle (-)}}{2},
\frac{\sigma^{\scriptscriptstyle (+)}}{2}$ and $\sigma_z$. The matrix 
$\displaystyle \frac{\sigma^{\scriptscriptstyle (-)}}{2}$ plays the role 
of the operator of multiplication by $c$. In fact:
\begin{eqnarray}
&& \pmatrix{0 & 0\cr 1 & 0}\cdot\left(\begin{array}{c} 1\\0\end{array}\right)=
\left(\begin{array}{c} 0\\1\end{array}\right)
\;\;\Longleftrightarrow\;\;\widehat{c}\cdot 1=c\nonumber\\
&& \pmatrix{0 & 0\cr 1 & 0}\cdot\left(\begin{array}{c} 0\\1\end{array}\right)=
\left(\begin{array}{c} 0\\0\end{array}\right)
\;\;\Longleftrightarrow\;\;\widehat{c}\cdot c=0.
\end{eqnarray}
The matrix $\displaystyle \frac{\sigma^{\scriptscriptstyle (+)}}{2}$ plays the role 
of the operator $\overline{c}$ which is the derivative operator with respect to $c$. In fact:
\begin{eqnarray}
\displaystyle
&& \pmatrix{0 & 1\cr 0 & 0}\left(\begin{array}{c} 1\\0\end{array}\right)=
\left(\begin{array}{c} 0\\0\end{array}\right)
\;\;\Longleftrightarrow\;\;\frac{\partial}{\partial c}1=0\nonumber\\
&& \pmatrix{0 & 1\cr 0 & 0}\left(\begin{array}{c} 0\\1\end{array}\right)=
\left(\begin{array}{c} 1\\0\end{array}\right)
\;\;\Longleftrightarrow\;\;\frac{\partial}{\partial c}c=1.
\end{eqnarray}
In the representation (\ref{5-1}) also the matrix $\sigma_z$ made its appearance and the reader 
may wonder on which is its role.
Actually the matrix $\sigma_z$ allows us to give to the states
$\left(\begin{array}{c} 1\\0\end{array}\right)$, $\left(\begin{array}{c} 0\\1\end{array}\right)$
the grading factors that 
one has to introduce for the anticommutativity
of the Grassmann variables. In fact the state 
$\left(\begin{array}{c} 1\\0\end{array}\right)$ must be Grassmannian even in order to represent ``1", 
while the state 
$\left(\begin{array}{c} 0\\1\end{array}\right)$ must be Grassmannian odd in order
to represent ``$c$". 
The matrix $\sigma_z$ gives exactly the correct grading factor to the vectors $\left(\begin{array}{c}
1\\0\end{array}\right)$ and $\left(\begin{array}{c} 0\\1\end{array}\right)$:
\begin{equation}
\displaystyle \sigma_z\left(\begin{array}{c} 1\\0\end{array}\right)=
\left(\begin{array}{c} 1\\0\end{array}\right), \;\;\;\;\;\;\;\sigma_z\left
(\begin{array}{c} 0\\1\end{array}\right)=-
\left(\begin{array}{c} 0\\1\end{array}\right).
\end{equation}
The presence of $\sigma_z$ in (\ref{5-1}) has the same goal. Let us in fact
give the representation of the equation 
$\displaystyle \widehat{\overline{c}}_q[c^pc^q]=\frac{\partial}{\partial c^q}(c^pc^q)=-c^p$ 
in terms of Pauli matrices in the case $n=1$. It is:
\begin{equation}
\displaystyle 
\sigma_z\otimes\frac{\sigma^{\scriptscriptstyle (+)}}{2}\Biggl[\left
(\begin{array}{c} 0\\1\end{array}\right)\otimes
\left(\begin{array}{c} 0\\1\end{array}\right)\Biggr]=\sigma_z\left(
\begin{array}{c} 0\\1\end{array}\right)
\otimes \frac{\sigma^{\scriptscriptstyle (+)}}{2}\left(\begin{array}{c} 
0\\1\end{array}\right)=-
\left(\begin{array}{c} 0\\1\end{array}\right)\otimes\left(\begin{array}
{c} 1\\0\end{array}\right). \label{5-14}
\end{equation}
Note that the matrix $\sigma_z$ is crucial in order to reproduce the minus sign 
on the RHS of Eq. (\ref{5-14}). That minus sign was there in the original equations written
in terms of $c$ and was due to the fact that the derivative
$\displaystyle \frac{\partial}{\partial c^q}$ had to go through a Grassmannian odd variable
$c^p$ in order to act on $c^q$. 

\section{Cartan Calculus and Pauli Matrices}

With the tools developed in the previous section, we can now generalize to more than one degree of freedom
what we did in Section {\bf 4}, that means to write down all the operations of the Cartan calculus via
Pauli matrices. 

\medskip

\noindent $\bullet$ {\bf Exterior Derivative}. The exterior
derivative ${\bf d}=\widehat{c}^a\partial_a$ is a linear operator
in the variables $c$. From (\ref{general}) its matrix representation is given by:
\begin{equation}
\displaystyle {\bf d}=\sum_{i=1}^{2n}=\widehat{c}^a\partial_a=
\sum_{j=1}^{2n}(\sigma_z)^{\otimes j-1}\otimes \frac{\sigma^{
\scriptscriptstyle (-)}}{2}\partial_j\otimes ({\bf 1})^{\otimes 
2n-j} \label{exteriorder}.
\end{equation} 

\medskip

\noindent $\bullet$ {\bf Form Number}. From the tensor expression (\ref{5-1})
of $\widehat{c}$ and $\widehat{\overline{c}}$ it is very easy to find the 
expression of $\widehat{Q}_f$ in the general case:
\begin{equation}
\displaystyle
\widehat{Q}_f=\sum_{j=1}^{2n}\widehat{c}^j\widehat{\overline{c}}_j=
\sum_{j=1}^{2n}(\sigma_z\cdot\sigma_z)
^{\otimes j-1}\otimes 
\frac{\sigma^{\scriptscriptstyle (-)}}{2}\cdot
\frac{\sigma^{\scriptscriptstyle (+)}}{2}\otimes ({\bf 1})^{\otimes 2n-j}.
\end{equation}
Since $\sigma_z\cdot\sigma_z={\bf 1}$ and $\displaystyle 
\frac{\sigma^{
\scriptscriptstyle (-)}}{2}\cdot
\frac{\sigma^{\scriptscriptstyle (+)}}{2}=\pmatrix{0 & 0\cr 0 & 1}=
\frac{1}{2}({\bf 1}-\sigma_z)$
we can rewrite $\widehat{Q}_f$ as:
\begin{equation}
\displaystyle
\widehat{Q}_f=\sum_{j=1}^{2n}\widehat{c}^j\widehat{\overline{c}}_j=
\sum_{j=1}^{2n}({\bf 1})^{\otimes j-1}
\otimes \frac{1}{2}
({\bf 1}-\sigma_z)\otimes ({\bf 1})^{\otimes 2n-j}.
\end{equation}
We call $\widehat{Q}_f$ the form number because 
an homogeneous form of degree $p$ 
is an eigenstate for the matrix $\widehat{Q}_f$ with eigenvalue $p$. 

\medskip

\noindent $\bullet$ {\bf Interior Contraction}. In the case $n=1$ 
the interior contraction with a vector
field was given by $\iota_{\scriptscriptstyle V}=V^q\widehat{\overline{c}}_q+V^p
\widehat{\overline{c}}_p$. In general we have:
$\iota_{\scriptscriptstyle V}=V^j\widehat{\overline{c}}_j$ whose matrix 
representation is given by:
\begin{equation}
\displaystyle \iota_{\scriptscriptstyle V}=\sum_{j=1}^{2n}(\sigma_z)^
{\otimes j-1}\otimes \frac{\sigma^{
\scriptscriptstyle (+)}}{2}V^j
\otimes ({\bf 1})^{\otimes 2n-j}.
\end{equation}
In the particular case of a Hamiltonian vector field $V^j=\omega^{jk}
\partial_kH$ we obtain:
\begin{equation}
\displaystyle \iota_{\scriptscriptstyle h}=\sum_{j,k=1}^{2n}(\sigma_z)^
{\otimes j-1}\otimes \frac{\sigma^{
\scriptscriptstyle (+)}}{2}\omega^{jk}
\partial_kH\otimes ({\bf 1})^{\otimes 2n-j}. \label{inthamvec}
\end{equation}

\medskip

\noindent $\bullet$ {\bf Lie Derivative along the Hamiltonian Flow}. 
It is easy to represent the Lie derivative along the Hamiltonian flow
as a matrix starting from the matrix representation of the exterior derivative {\bf d}, 
Eq. (\ref{exteriorder}), and of the interior contraction with a Hamiltonian vector field,
Eq. (\ref{inthamvec}). In fact, remembering \cite{Marsden} that the Lie derivative along the Hamiltonian
vector field is given by the anticommutator of ${\bf d}$ and $\iota_h$: 
${\cal L}_h={\bf d}\iota_h+\iota_h{\bf d}$ we have that:
\begin{eqnarray}
\displaystyle {\cal L}_h={\bf d}\iota_h+\iota_h{\bf d}&=&\sum_{j<k}({\bf 1})^{\otimes j-1}
\otimes \biggl[\sigma_z,\frac{\sigma^{\scriptscriptstyle (+)}}{2}\biggr]\otimes (\sigma_z)^{\otimes k-j-1}
\otimes \frac{\sigma^{\scriptscriptstyle (-)}}{2}\otimes ({\bf 1})^{\otimes 2n-k}\omega^{jl}\partial_lH
\partial_k\nonumber\\
& &+\sum_{j<k}({\bf 1})^{\otimes j-1}
\otimes \sigma_z\cdot\frac{\sigma^{\scriptscriptstyle (+)}}{2}\otimes (\sigma_z)^{\otimes k-j-1}
\otimes \frac{\sigma^{\scriptscriptstyle (-)}}{2}\otimes ({\bf 1})^{\otimes 2n-k}\omega^{jl}\partial_k\partial_lH
\nonumber\\
& &+\sum_j({\bf 1})^{\otimes j-1}
\otimes {\bf 1} \otimes ({\bf 1})^{\otimes 2n-j}\omega^{jl}\partial_lH\partial_j+ \label{6-6}
\\
& &\sum_{j>k}({\bf 1})^{\otimes k-1}
\otimes \biggl[\frac{\sigma^{\scriptscriptstyle (-)}}{2},\sigma_z\biggr]\otimes (\sigma_z)^{\otimes j-k-1}
\otimes \frac{\sigma^{\scriptscriptstyle (+)}}{2}\otimes ({\bf 1})^{\otimes 2n-j}\omega^{jl}\partial_lH\partial_k
\nonumber\\
& &+\sum_{j>k}({\bf 1})^{\otimes k-1}
\otimes \frac{\sigma^{\scriptscriptstyle (-)}}{2}\cdot\sigma_z\otimes (\sigma_z)^{\otimes j-k-1}
\otimes \frac{\sigma^{\scriptscriptstyle (+)}}{2}\otimes ({\bf 1})^{\otimes 2n-j}\omega^{jl}\partial_k\partial_lH
\nonumber
\end{eqnarray}
Using the anticommutation relations $\displaystyle \biggl[\sigma_z,\frac{\sigma^{\scriptscriptstyle (+)}}{2}
\biggr]_{\scriptscriptstyle +}=\biggl[\frac{\sigma^{\scriptscriptstyle (-)}}{2},\sigma_z
\biggr]_{\scriptscriptstyle +}=0$ we can re-write (\ref{6-6}) in the more compact form:
\begin{eqnarray}
\label{seiotto} 
{\cal L}_h=&&(\omega^{ab}\partial_bH\partial_a)({\bf 1})
^{\otimes 2n}+\\
&&-\sum_{j<k}({\bf 1})^{\otimes j-1}\otimes \frac{\sigma^
{\scriptscriptstyle (+)}}{2}
\sigma_z\otimes (\sigma_z)^{\otimes k-1-j}\otimes 
\frac{\sigma^{\scriptscriptstyle (-)}}{2}\otimes 
({\bf 1})^{\otimes 2n-k}\cdot \omega^{jl}\partial_l\partial_kH+\nonumber\\
&&-\sum_{j>k}({\bf 1})^{\otimes k-1}\otimes \sigma_z\frac{\sigma^{\scriptscriptstyle
(-)}}{2}
\otimes (\sigma_z)^{\otimes j-1-k}\otimes \frac{\sigma^{\scriptscriptstyle 
(+)}}{2}
\otimes ({\bf 1})^{\otimes 2n-j}\cdot \omega^{jl}\partial_l\partial_kH. \nonumber 
\end{eqnarray}
It is easy to realize that Eq. (\ref{seiotto}) is just $i$ times the matrix representation of
$\HT$:
\begin{equation}
{\cal L}_h=i\widehat{\HT}=\omega^{ab}\partial_bH\partial_a-
\widehat{\overline{c}}_j\omega^{jl}\partial_l\partial_kH\widehat{c}^k.
\end{equation}
To check that, we just have to use in $\widehat{\cal H}$ the usual matrix representation 
(\ref{5-1}) for the fermionic operators $\widehat{c},\widehat{\overline{c}}$:
\begin{eqnarray}
\displaystyle 
&&\widehat{c}^k=(\sigma_z)^{\otimes k-1}\otimes \frac{\sigma^
{\scriptscriptstyle (-)}}{2}\otimes 
({\bf 1})^{\otimes 2n-k},\nonumber\\
&&\widehat{\overline{c}}_j=(\sigma_z)^{\otimes j-1}\otimes \frac{\sigma^
{\scriptscriptstyle 
(+)}}{2}\otimes ({\bf 1})^{\otimes 2n-j}.
\end{eqnarray}
This confirms that the operator $\HT$ which appears in the weight of the CPI (\ref{CPI}) is nothing else than 
the Lie derivative along the Hamiltonian flow.
\medskip

\noindent $\bullet$ {\bf The Adjoint of {\bf d} and the Laplacian}. 
The Grassmann expression of $\delta$ is $\delta=-\overline{c}_j\partial_j$ and using (\ref{5-1})
we obtain that in terms of Pauli matrices:
\begin{equation}
\displaystyle
\delta=-\widehat{\overline{c}}_j\partial_j=-\sum_{j=1}^{2n}(\sigma_z)^
{\otimes j-1}\otimes \frac{\sigma^{
\scriptscriptstyle (+)}}{2}\partial_j
\otimes ({\bf 1})^{\otimes 2n-j}.
\end{equation}
This is the expression of $\delta$ for an arbitrary number of degrees of freedom $n$. It 
is possible to prove that ${\bf d}$ and $\delta$
are nilpotent. Let us check that for the exterior derivative ${\bf d}$:
\begin{eqnarray}
\displaystyle
{\bf d}^2 &=&\sum_{jk}\biggl[(\sigma_z)^{\otimes j-1}\otimes \frac{\sigma^
{\scriptscriptstyle 
(-)}}{2}\partial_j\otimes ({\bf 1})^{\otimes 2n-j}\biggr]
\biggl[(\sigma_z)^{\otimes k-1}\otimes \frac{\sigma^{\scriptscriptstyle 
(-)}}{2}\partial_k\otimes ({\bf 1})^{\otimes 2n-k}\biggr]=\nonumber\\
&=&\sum_{j<k}({\bf 1})^{\otimes j-1}\otimes\frac{\sigma^{\scriptscriptstyle 
(-)}}{2}\sigma_z\partial_j\otimes (\sigma_z)^{\otimes k-1-j}
\otimes \frac{\sigma^{\scriptscriptstyle (-)}}{2}\partial_k\otimes 
({\bf 1})^{\otimes 2n-k}+\nonumber\\
& &+\sum_{k<j}({\bf 1})^{\otimes k-1}\otimes \sigma_z\frac{\sigma^
{\scriptscriptstyle 
(-)}}{2}\partial_k\otimes (\sigma_z)^{\otimes j-1-k}
\otimes \frac{\sigma^{\scriptscriptstyle (-)}}{2}\partial_j\otimes 
({\bf 1})^{\otimes 2n-j}=\nonumber\\
&=&\sum_{j<k}({\bf 1})^{\otimes j-1}\otimes \biggl[\frac{\sigma^
{\scriptscriptstyle 
(-)}}{2},\sigma_z\biggr]_+\partial_j\otimes 
(\sigma_z)^{\otimes k-1-j}\otimes \frac{\sigma^{\scriptscriptstyle (-)}}
{2}\partial_k\otimes ({\bf 1})^{\otimes 2n-k}=\nonumber\\
&=&0
\end{eqnarray}
where we have used respectively the fact that:\newline\smallskip
{\bf 1)} the terms with $j=k$ do not contribute to the sum since 
$\displaystyle \frac{\sigma^{\scriptscriptstyle (-)}}{2}$ is nilpotent;\newline \smallskip
{\bf 2)} $j$ and $k$ are dummy indices and so they can be interchanged; \newline \smallskip
{\bf 3)} $\displaystyle\biggl[\frac{\sigma^{\scriptscriptstyle (-)}}{2},\sigma_z\biggr]_+=0$. \newline \smallskip
With an analogous 
calculation one can prove that $\delta^2=0$.
Because of this the Laplacian turns out to be just the anticommutator of ${\bf d}$ and $\delta$:
$\Delta=({\bf d}+\delta)^2=[{\bf d},\delta]_+$. Using this and the 
matrix representations of ${\bf d}$ and $\delta$
we have for the Laplacian the following expression:
\begin{eqnarray}
\displaystyle \Delta &=&-\biggl[\sum_j(\sigma_z)^{\otimes j-1}\otimes \frac{\sigma^{
\scriptscriptstyle (-)}}
{2}\partial_j\otimes ({\bf 1})^{\otimes 2n-j},
\sum_k(\sigma_z)^{\otimes k-1}\otimes \frac{\sigma^{
\scriptscriptstyle (+)}}{2}\partial_k
\otimes ({\bf 1})^{\otimes 2n-k}\biggr]_+=\nonumber\\
&=&-\sum_{j<k}({\bf 1})^{\otimes j-1}\otimes \biggl[\frac{\sigma^{\scriptscriptstyle (-)}}{2},
\sigma_z\biggr]_+\partial_j
\otimes (\sigma_z)^{\otimes k-j-1}\otimes \frac{\sigma^{\scriptscriptstyle (+)}}{2}
\partial_k\otimes({\bf 1})^{\otimes 2n-k}\nonumber\\
& &-\sum_j({\bf 1})^{\otimes j-1}\otimes\biggl[\frac{\sigma^{\scriptscriptstyle (-)}}{2},
\frac{\sigma^{\scriptscriptstyle (+)}}{2}\biggr]_+
\partial_j^2\otimes ({\bf 1})^{\otimes 2n-j}\nonumber\\
& &-\sum_{j>k}({\bf 1})^{\otimes k-1}\otimes\biggl[\sigma_z,
\frac{\sigma^{\scriptscriptstyle (+)}}{2}\biggr]_+\partial_k
\otimes (\sigma_z)^{\otimes j-k-1}\otimes\frac{\sigma^{\scriptscriptstyle (-)}}{2}
\partial_j\otimes ({\bf 1})^{\otimes 2n-j}=\nonumber\\
&=&-\sum_j({\bf 1})^{\otimes j-1}\otimes {\bf 1}\partial_j^2
\otimes ({\bf 1})^{\otimes 2n-j}.
\end{eqnarray}
In the notations of the CPI we have for $\Delta$ the following 
positive definite operator:
\begin{equation}
\displaystyle
\Delta=\sum_{j=1}^{2n}({\bf 1})^{\otimes j-1}\otimes {\bf 1}
\lambda_j^2\otimes ({\bf 1})^{\otimes 2n-j}.
\end{equation}

\medskip

\noindent $\bullet$ {\bf Other Symmetry Charges of the CPI}. 
In the CPI some other charges were found \cite{Gozzi} which had also a clear geometrical meaning.
For completeness we will write down here their expression in terms of Pauli matrices:
\begin{eqnarray}
\displaystyle
&&\widehat{\overline{Q}}=\widehat{\overline{c}}_a\omega^{ab}\partial_b=
\sum_{j,l=1}^{2n}(\sigma_z)^{\otimes j-1}
\otimes \frac{\sigma^{\scriptscriptstyle (+)}}{2}\omega^{jl}\partial_l
\otimes ({\bf 1})^{\otimes 2n-j}\nonumber\\
&&\widehat{K}=\frac{1}{2}\omega_{ab}\widehat{c}^a\widehat{c}^b=\sum_{i=1}^n
\widehat{c}^{p_i}\widehat{c}^{q_i}=\nonumber\\
&&\qquad\;=\sum_{i=1}^n\biggl[(\sigma_z)^{\otimes 2(i-1)}\otimes
\frac{\sigma^{\scriptscriptstyle (-)}}{2}
\otimes ({\bf 1})^{\otimes 2(n-i)+1}\biggr]
\biggl[(\sigma_z)^{\otimes 2i-1}\otimes\frac{\sigma^{\scriptscriptstyle 
(-)}}{2}\otimes ({\bf 1})^{\otimes 2(n-i)}\biggr]=\nonumber\\
&&\qquad\;=\sum_{i=1}^n({\bf 1})^{\otimes 2(i-1)}\otimes 
\frac{\sigma^{\scriptscriptstyle 
(-)}}{2}\sigma_z\otimes\frac{\sigma^{\scriptscriptstyle (-)}}{2}\otimes
({\bf 1})^{\otimes 2(n-i)}=\nonumber\\
&&\qquad\;=\sum_{i=1}^n({\bf 1})^{\otimes 2i-2}\otimes 
\biggl(\frac{\sigma^{\scriptscriptstyle 
(-)}}{2}\biggr)^{\otimes 2}\otimes({\bf 1})^{\otimes 2n-2i}\nonumber\\
&&\widehat{\overline{K}}=\sum_{i=1}^{n}\widehat{\overline{c}}_{q_i}\widehat{\overline{c}}_{p_i}
=\sum_{i=1}^n({\bf 1})^{\otimes 2i-2}
\otimes\biggl(\frac{\sigma^{\scriptscriptstyle 
(+)}}{2}\biggr)^{\otimes 2}\otimes({\bf 1})^{\otimes
2n-2i} \\
&&\widehat{Q}_{\scriptscriptstyle H}=\sum_{j=1}^{2n}\widehat{c}^j(\partial_j-\beta\partial_jH)=
\sum_{j=1}^{2n}(\sigma_z)
^{\otimes j-1}\otimes\frac{\sigma^{\scriptscriptstyle
(-)}}{2}(\partial_j-\beta\partial_jH)
\otimes({\bf 1})^{\otimes 2n-j}\nonumber\\
&&\widehat{\overline{Q}}_{\scriptscriptstyle H}=\sum_{j,l=1}^{2n}\widehat{\overline{c}}_j\omega^{jl}
(\partial_l+\beta\partial_lH)=\sum_{j,l=1}^{2n}
(\sigma_z)^{\otimes j-1}\otimes\frac{\sigma^{\scriptscriptstyle
(+)}}{2}\omega^{jl}
(\partial_l+\beta\partial_lH)\otimes ({\bf 1})^{\otimes 2n-j}. \nonumber
\end{eqnarray}
\section{Summary and Conclusions}
In this paper we have shown that the differential forms on a symplectic manifold can be mapped into suitable
tensor products of 2-dimensional vectors $\left(\begin{array}{c} 1\\0 \end{array} \right)$ and
$\left(\begin{array}{c} 0\\1 \end{array} \right)$. Similarly all the operations of the Cartan calculus
like exterior derivatives, interior contractions, Lie derivatives can be represented by suitable
tensor products of Pauli matrices. This sort of mapping was made easy by the results of Ref. \cite{Gozzi} 
where all the abstract concepts of differential geometry on symplectic manifolds were turned 
into operations on Grassmannian variables. The next step was to give a matrix representation to these Grassmannian
variables. 

The reader may ask now two questions. The first is if these results can be generalized to arbitrary 
non-symplectic manifolds. We feel this is possible by first going through a sort of formalism like that of Ref.
\cite{Gozzi}. Basically we will first give a functional representation of the vector flows on this manifold
as in \cite{Gozzi} and then study the geometrical meaning of the Grassmannian variables which will naturally
appear. Second we will try to represent these Grassmannian variables and operators via vectors and matrices. 
In the case of symplectic manifolds these matrices are the tensor product of an {\it even} number of Pauli matrices,
in the general case they should reduce to tensor
products of an {\it arbitrary} number of Pauli matrices. 
The second question the reader may ask is which is the usefulness of what we did in this paper. {\it First} we think
that, having reduced the differential forms to vectors and the 
Cartan operations to tensor products of Pauli matrices, may help 
in building computer packages to do differential calculus. {\it Second}, as the variables $c$ could be interpreted not only
as forms but also as Jacobi fields \cite{Gozzi} and a lot of information on concepts like ergodicity and chaos
\cite{Liapunov}\cite{Deotto}  could be extracted from them, we think that their new representation as tensor
products of 2-dimensional vectors could help in having simpler computer simulations of these systems.
{\it Last}, but not least, a further application of this formalism could be in the field of ``quantum computation" 
\cite{nielsen}. In this field a crucial role is taken by the ``spin" 1/2 variables needed in order to implement logical 
gates. In classical mechanics (CM) no spin 1/2 variable seemed possible. In this paper instead we have proved 
that if we include differential forms in CM, there is a natural appearance of concepts like Pauli matrices and spin 1/2
states. Moreover, besides this, CM can be endowed with a Hilbert space \cite{dgm} structure, 
so this formulation seems to be the perfect one in which to handle controversial issues, nowadays present in the field of 
quantum computation \cite{nielsen}, of how much some features of quantum computation 
are truly quantum and how much they are instead just classical. 

\section*{Acknowledgments}

I wish to thank G. Pastore for some questions which triggered the present investigation.
A very special thank to E. Gozzi
for a lot of useful discussions and help and to M.V. Ioffe for some important technical suggestions. This work 
has been supported in part by funds from INFN, MIUR and the University of Trieste.

\appendix
\makeatletter
\@addtoreset{equation}{section}
\makeatother
\renewcommand{\theequation}{\thesection.\arabic{equation}}

\section{Appendix  }
{\centerline{\bf Evolution of a Generalized Wave Function in the Matrix Formulation}}

In this Appendix we want to give a very simple example of how it is possible to reconstruct
the evolution of a generalized wave function in the matrix formulation of the CPI.
Suppose we consider the generic wave function:
\begin{equation}
\psi(t=0)=\psi_{\scriptscriptstyle 0}(\varphi)\left(\begin{array}{c} 1\\0\\0\\0 \end{array} \right)
+\psi_q(\varphi)\left(\begin{array}{c} 0\\1\\0\\0 \end{array} \right)\nonumber\\
+\psi_p(\varphi)\left(\begin{array}{c} 0\\0\\1\\0 \end{array} \right)+
\psi_{\scriptscriptstyle 2}(\varphi)\left(\begin{array}{c} 0\\0\\0\\1 \end{array} \right).
\end{equation}
We want to reconstruct its evolution in the free particle case, i.e. with 
$H=p^2/2$. Inserting this $H$ into the $\widehat{\cal H}$ of
(\ref{accatilde}) we get:
\begin{equation}
\widehat{\cal H}=\widehat{L}\cdot {\bf 1}_{4\times 4}+i\widehat{\overline{c}}_q\cdot\widehat{c}^p=
\pmatrix{\widehat{L} & 0 & 0 & 0\cr 0 & \widehat{L} & 0 & 0\cr 0 & -i & \widehat{L} & 0\cr 0 & 0 &
0 & \widehat{L}}.
\end{equation}
The wave function at time $t$ will be given by:
\begin{eqnarray}
&&\label{funt} \displaystyle \psi(t)=e^{-i\widehat{\cal H}t}\psi(t=0)=e^{-i\widehat{L}t}e^{\widehat{\overline{c}}_q \widehat{c}^pt}
\psi(t=0)=e^{-i\widehat{L}t}(1+\widehat{\overline{c}}_q\widehat{c}^pt)\psi(t=0)=\nonumber\\
&& =e^{-i\widehat{L}t}\psi_{\scriptscriptstyle 0}(\varphi)\pmatrix{1 & 0 & 0 & 0\cr
0 & 1 & 0 & 0\cr 0 & -t & 1 & 0\cr 0 & 0 & 0 & 1}\left(\begin{array}{c} 1\\0\\0\\0 \end{array} \right)
+e^{-i\widehat{L}t}\psi_{q}(\varphi)\pmatrix{1 & 0 & 0 & 0\cr
0 & 1 & 0 & 0\cr 0 & -t & 1 & 0\cr 0 & 0 & 0 & 1}\left(\begin{array}{c} 0\\1\\0\\0 \end{array}\right)
\nonumber\\
&& +e^{-i\widehat{L}t}\psi_{p}(\varphi)\pmatrix{1 & 0 & 0 & 0\cr
0 & 1 & 0 & 0\cr 0 & -t & 1 & 0\cr 0 & 0 & 0 & 1}\left(\begin{array}{c} 0\\0\\1\\0 \end{array} \right)
+e^{-i\widehat{L}t}\psi_{\scriptscriptstyle 2}(\varphi)\pmatrix{1 & 0 & 0 & 0\cr
0 & 1 & 0 & 0\cr 0 & -t & 1 & 0\cr 0 & 0 & 0 & 1}\left(\begin{array}{c} 0\\0\\0\\1 \end{array} \right).
\nonumber\\
&&
\end{eqnarray}
We know that the evolution of a function $\psi_i(\varphi)$ with the Liouvillian does not alter its functional form
but forces us only to replace its argument $\varphi$ with $\varphi_t=(q_t,p_t)$ that, in the case of a free particle, 
is given by $\varphi_t=(q-pt,p)$, see \cite{waves}.
So, if we identify with $\psi_i(\varphi_t)$ the evolution with the Liouvillian of the $i$-th component 
of the wave function:
\begin{equation}
\displaystyle \psi_i(\varphi_t)=e^{-i\widehat{L}t}\psi_i(\varphi)
\end{equation}
we can rewrite the wave function (\ref{funt}) in a compact way as:
\begin{equation}
\psi(t)=\left( \begin{array}{c} \psi_{\scriptscriptstyle 0}(\varphi_t) \\ \psi_q(\varphi_t)\\ \psi_p(\varphi_t)-t\psi_q(\varphi_t)
\\ \psi_{\scriptscriptstyle 2}(\varphi_t) \end{array}\right).
\end{equation}
According to the conventions of Eq. (\ref{genw2}) we have that the expansion of $\psi(\varphi,c)$ in terms of the $c$'s
is given by 
\begin{eqnarray}
\psi(t)&=&\psi_{\scriptscriptstyle 0}(\varphi_t)+\psi_q(\varphi_t)c^q+
\bigl(\psi_p(\varphi_t)-t\psi_q(\varphi_t)\bigr)c^p+\psi_{\scriptscriptstyle 2}(\varphi_t)
c^pc^q=\nonumber\\
&=&\psi_{\scriptscriptstyle 0}(\varphi_t)+\psi_q(\varphi_t)(c^q-tc^p)+\psi_p(\varphi_t)c^p+
\psi_{\scriptscriptstyle 2}(\varphi_t)c^pc^q.
\label{beta}
\end{eqnarray}
The same result can be obtained by means of the kernel of propagation of the CPI \cite{Gozzi}:
\begin{equation}
K(\varphi,c,t|\varphi_i,c_i,t)=\delta(\varphi(t)-\varphi_{cl}(t))\delta(c(t)-c_{cl}(t)). \label{ker}
\end{equation}
In fact in the case of a free particle (\ref{ker}) becomes:
\begin{equation}
K(\varphi,c,t|\varphi_i,c_i,t_i)=\delta(q-q_i-p_it)\delta(p-p_i)\delta(c^q-c^q_i-c^p_it)\delta(c^p-c^p_i)
\end{equation}
and the wave function at time $t$ is
\begin{eqnarray}
\psi(\varphi,c,t)&=&\int dq_idp_idc^p_idc^q_i
K(\varphi,c,t|\varphi_i,c_i,t_i)\psi(\varphi_i,c_i)=\nonumber\\
&=&\psi(q-pt,p,c^q-tc^p,c^p,0). \label{gamma}
\end{eqnarray}
The RHS of Eq. (\ref{gamma}) coincides exactly with the result obtained in Eq. (\ref{beta})
where $\varphi_t=(q-pt,p)$.

\section{Appendix }
{\centerline{\bf The Group of Symmetry Charges of the CPI}}

It is well-known, see paper \cite{Gozzi}, that the charges $Q_f$, $K$ and $\overline{K}$ 
make an algebra that is, modulus a central extension, the algebra of the
group Sp(2):
\begin{eqnarray}
&&\label{spp2} [Q_f,K]_{\scriptscriptstyle -}=2K\nonumber\\
&&[Q_f,\overline{K}]_{\scriptscriptstyle -}=-2\overline{K}\\
&&[K,\overline{K}]_{\scriptscriptstyle -}=Q_f-1. \nonumber 
\end{eqnarray}
It is easy to check that the $4\times 4$ matrices $\widehat{K},\widehat{\overline{K}}, \widehat{Q}_f$ 
defined in (\ref{opmat}) satisfy (\ref{spp2}).
Since the matrix ${\bf 1}_{\scriptscriptstyle 4\times 4}$ commutes with 
$\widehat{K}$ and $\widehat{\overline{K}}$ 
it is also possible to throw away 
the central extension of the algebra by replacing $\widehat{Q}_f\to 
\widehat{Q}_f-{\bf 1}_{\scriptscriptstyle 4\times 4}$.
What we want to prove now is that $2\times 2$ matrices are sufficient 
to reproduce the Sp(2) algebra (\ref{spp2}). 
Not only, but the matrices we have to 
consider are just the
matrices $\displaystyle \frac{\sigma^{\scriptscriptstyle (+)}}{2},
\frac{\sigma^{\scriptscriptstyle (-)}}{2},\sigma_z$ 
we used several times in the previous sections.
In fact:
\begin{eqnarray}
\displaystyle 
&&\biggl[\sigma_z,\frac{\sigma^{\scriptscriptstyle (+)}}{2}\biggr]_{\scriptscriptstyle -}=
\pmatrix{0 & 2\cr 0 & 0}=2\biggl(\frac{\sigma^{(+)}}{2}\biggr)\nonumber\\
&&\biggl[\sigma_z,\frac{\sigma^{\scriptscriptstyle (-)}}{2}\biggr]_{\scriptscriptstyle -}=
\pmatrix{0 & 0\cr -2 & 0}=-2\biggl(\frac{\sigma^{\scriptscriptstyle (-)}}{2}\biggr)\\
&&\biggl[\frac{\sigma^{\scriptscriptstyle (+)}}{2},\frac{\sigma^{\scriptscriptstyle (-)}}{2}\biggr]_{\scriptscriptstyle -}=
\pmatrix{1 & 0\cr 0 & -1}=\sigma_z. \nonumber
\end{eqnarray} 
So, if we identify $\displaystyle Q_f=\sigma_z,\; K
=\frac{\sigma^{\scriptscriptstyle (+)}}{2}, \; 
\overline{K}=\frac{\sigma^{\scriptscriptstyle (-)}}{2}$,
we can reproduce
the Sp(2) algebra, without any central extension, in terms of Pauli matrices. 
Moreover the only $2\times 2$ matrices that commute with the 3 generators of 
Sp(2) are the matrices proportional to the identity.
This fact confirms that the representation we found is irreducible.
We note however that, while it is possible to represent the 
generators of Sp(2) as $2\times 2$ matrices,
it is completely impossible to include also the $Q$, 
$\overline{Q}$ charges. 
For example it is impossible to find a $2\times 2$
matrix $Q$ which satisfies the relation 
$[Q_f,Q]_{\scriptscriptstyle -}=Q$. 
Therefore if we want to extend Sp(2) including all the other symmetry 
charges of the CPI we have to consider $4\times 4$ instead of $2\times 2$
matrices in order to find non trivial representations of the algebra. 
For sure the matrices of operators (\ref{opmat})-(\ref{matrixn1}) satisfy the correct algebra
of the symmetry charges of the CPI. What we want to prove now is that it is possible 
to find an irreducible representation of the symmetry charges of the CPI 
in terms of $4\times 4$ matrices whose entries are {\it real numbers}.
Suppose we consider, as independent charges, $Q=ic^a\lambda_a,\; 
\overline{Q}=i\overline{c}_a\omega^{ab}\lambda_b,
\; -i\overline{N}=-i\overline{c}_a\omega^{ab}\partial_bH,\; iN=ic^a\partial_aH$. 
They are 4 conserved charges and the 
only anticommutators different from zero are: $[Q,-i\overline{N}]_{\scriptscriptstyle +}=
[\overline{Q},iN]_{\scriptscriptstyle +}=\HT$. Since $\HT$ is a Casimir for
the entire algebra the irreducible representations will be labeled by its 
eigenvalues $h$. Using
the Appendix of Ref. \cite{Rittenberg} about the irreducible 
representations of S(2), we can start by considering
two basis vectors $e_1,e_2$ to represent the subalgebra $[Q,-i\overline{N}]_{\scriptscriptstyle +}=\HT$ 
and two basis vectors $f_1,f_2$
to represent the subalgebra $[\overline{Q},iN]_{\scriptscriptstyle +}=\HT$:
\begin{eqnarray}
&&Qe_1=\sqrt{h}e_2,\;\;\; -i\overline{N}e_1=0,\;\;\;\qquad\quad iNf_1=
\sqrt{h}f_2,\;\;\;\overline{Q}f_1=0,\nonumber\\
&&Qe_2=0,\;\;\; -i\overline{N}e_2=\sqrt{h}e_1,\;\;\;\qquad \quad iNf_2=0,
\;\;\;\overline{Q}f_2=\sqrt{h}f_1
\end{eqnarray}
where we choose $e_1, f_1$ as Grassmannian even and consequently 
$e_2, f_2$ as Grassmannian odd. A basis to represent the
algebra of $Q,\overline{Q},iN,-i\overline{N}, \HT$ is given by:
$F_1=e_1f_1$, $F_2=e_1f_2$, $F_3=e_2f_1$, $F_4=e_2f_2$.
Obviously $F_1$ and $F_4$ are Grassmannian even while $F_2$ and $F_3$ 
are Grassmannian odd. The matrix
representation of $Q$ is given by:
\begin{eqnarray}
&&QF_1=(Qe_1)f_1=\sqrt{h}e_2f_1=\sqrt{h}F_3\nonumber\\
&&QF_2=(Qe_1)f_2=\sqrt{h}e_2f_2=\sqrt{h}F_4\nonumber\\
&&\qquad\qquad\qquad\qquad\quad\Downarrow\nonumber\\
&&\qquad\quad Q=\pmatrix{0 & 0 & 0 & 0\cr 0 & 0 & 0 & 0\cr \sqrt{h} 
& 0 & 0 & 0\cr 0 & \sqrt{h} & 0 & 0}. \label{qmat}
\end{eqnarray}
In the same way we can find the matrices associated to the other 
symmetry charges:
\begin{eqnarray}
&&-i\overline{N}=\pmatrix{0 & 0 & \sqrt{h} & 0\cr 0 & 0 & 0 & \sqrt{h}\cr 
0 & 0 & 0 & 0\cr 0 & 0 & 0 & 0},\;\;\;\;
\overline{Q}=\pmatrix{0 & \sqrt{h} & 0 & 0\cr 0 & 0 & 0 & 0\cr 0 & 0 & 
0 & -\sqrt{h}\cr 0 & 0 & 0 & 0}\nonumber\\
&&iN=\pmatrix{0 & 0 & 0 & 0\cr \sqrt{h} & 0 & 0 & 0\cr 0 & 0 & 0 & 0\cr 
0 & 0 & -\sqrt{h} & 0},\;\;\;\;
\HT=\pmatrix{h & 0 & 0 & 0\cr 0 & h & 0 & 0\cr 0 & 0 & h & 0\cr 0 & 0 & 
0 & h}. \label{nummat}
\end{eqnarray}

Among the charges (\ref{qmat})-(\ref{nummat}) only the following anticommutators
are different from zero: $[Q,-i\overline{N}]_{\scriptscriptstyle +}=
[\overline{Q},iN]_{\scriptscriptstyle +}=\cal H$.
Among these $4\times 4$ matrices there is enough space also for the 
charges $Q_f=c^a\overline{c}_a$, $\displaystyle K=\frac{1}{2}\omega_{ab}c^ac^b$ 
and $\displaystyle \overline{K}=\frac{1}{2}\omega^{ab}\overline{c}_a\overline{c}_b$:
\begin{eqnarray}
&&\qquad \qquad Q_f=\pmatrix{0 & 0 & 0 & 0\cr 0 & 1 & 0 & 0\cr 0 & 0 & 1 & 0\cr 0 & 
0 & 0 & 2},\nonumber\\
&& K=\pmatrix{0 & 0 & 0 & 0\cr 0 & 0 & 0 & 0\cr 0 & 0 & 0 & 0\cr 1 & 0 & 
0 & 0},\;\;\;\;\;\;
\overline{K}=\pmatrix{0 & 0 & 0 & 1\cr 0 & 0 & 0 & 0\cr 0 & 0 & 0 & 
0\cr 0 & 0 & 0 & 0}. \label{sp2}
\end{eqnarray}
In Eq. (\ref{sp2}) we have found again the matrices of Eq. (\ref{opmat}).
The novelty is entirely contained in Eqs. (\ref{qmat})-(\ref{nummat})
since there we have matrices whose entries are real numbers instead of operators.
All the matrices we have defined here satisfy the correct algebra 
of the symmetry charges of the CPI, which 
can be found in the original papers \cite{Gozzi} or in Eq. (\ref{algcpi}).
It is possible to prove that the only $4\times 4$ matrix that commutes
with all the symmetry charges of the theory is given, modulus a proportionality 
factor, by the identity matrix. Therefore the representation we have found
is irreducible.  
So, in order to construct a non trivial irreducible representation of the symmetry charges of the CPI,
we just need 2 Grassmannian even states ($F_1$ and $F_4$) and 2 Grassmannian odd states ($F_2$ and $F_3$). 
All these 4 states can be connected each other by means of the symmetry charges as it emerges from Figure 1.
For example the charges $Q$ and $N$ which increase the form number by one allow us to go from $F_1$ 
to $F_2,F_3$ and from $F_2,F_3$ to $F_4$. The charge $K$ which increases the form number by 2 allows us to go from 
$F_1$ to $F_4$. In the opposite direction we can go via $\overline{K}$. 

\begin{figure}
\centering
\includegraphics{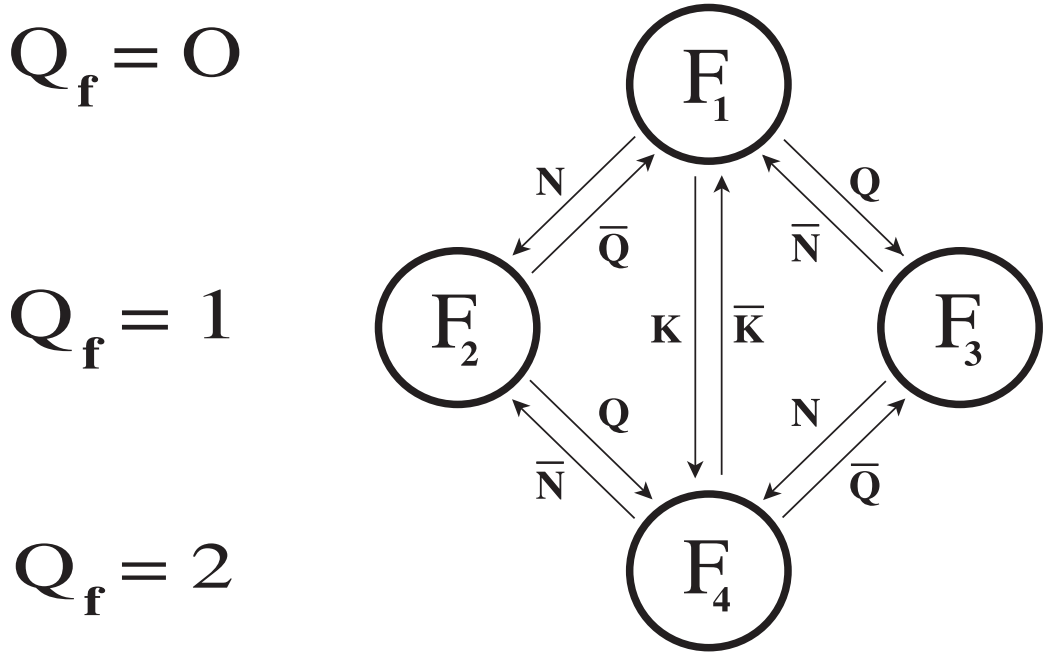}
\caption{Representation of the CPI charges} \label{rep}
\end{figure}

\section{Appendix }
{\centerline{\bf Calculational Details}

In this Appendix we will give the details of the derivation of formula (\ref{5-3}), i.e. we want to derive
the Grassmann algebra from the properties of the Pauli matrices. First of all we want to prove that all the 
$\widehat{c}$ anticommute.
If we take two indices $k$ and $l$ with $k<l$ we have that:
\begin{equation}
\displaystyle \widehat{c}^k\widehat{c}^l=
{\bf 1}\otimes {\bf 1}\otimes\ldots
\otimes {\bf 1}\otimes \underbrace{\frac{\sigma^{\scriptscriptstyle (-)}}{2}\sigma_z}_{k}
\otimes \sigma_z\otimes\sigma_z\otimes
\ldots\otimes \underbrace{\frac{\sigma^{\scriptscriptstyle (-)}}{2}}_{l}\otimes {\bf 1}\otimes\dots\otimes {\bf 1}
\end{equation}
while
\begin{equation}
\displaystyle \widehat{c}^l\widehat{c}^k=
{\bf 1}\otimes {\bf 1}\otimes\ldots
\otimes {\bf 1}\otimes \underbrace{\sigma_z\frac{\sigma^{\scriptscriptstyle (-)}}{2}}_{k}
\otimes \sigma_z\otimes\sigma_z\otimes
\ldots\otimes \underbrace{\frac{\sigma^{\scriptscriptstyle (-)}}{2}}_{l}\otimes {\bf 1}\otimes\dots\otimes {\bf 1}.
\end{equation}
Therefore the anticommutator is given by:
\begin{equation}
\displaystyle [\widehat{c}^k,\widehat{c}^l]_{\scriptscriptstyle +}=
{\bf 1}\otimes {\bf 1}\otimes\ldots
\otimes {\bf 1}\otimes \underbrace{\biggl[\frac{\sigma^{\scriptscriptstyle (-)}}{2},\sigma_z
\biggr]_{\scriptscriptstyle +}}_{k}
\otimes \sigma_z\otimes\sigma_z\otimes
\ldots\otimes \underbrace{\frac{\sigma^{\scriptscriptstyle (-)}}{2}}_{l}\otimes {\bf 1}\otimes\dots\otimes {\bf 1}
\end{equation}
and so $\widehat{c}^k$ and $\widehat{c}^l$ anticommute because $\displaystyle 
\biggl[\frac{\sigma^{\scriptscriptstyle (-)}}{2},\sigma_z\biggr]_{\scriptscriptstyle +}=0$. If instead we take $k=l$ 
we have immediately:
\begin{equation}
\displaystyle \widehat{c}^k\widehat{c}^k={\bf 1}\otimes {\bf 1}\otimes
\ldots\otimes{\bf 1}\otimes\biggl(\frac{\sigma^{\scriptscriptstyle (-)}}{2}\biggr)^2\otimes {\bf 1}\otimes
\ldots \otimes {\bf 1}.
\end{equation}
So $\widehat{c}^k \widehat{c}^k=0$ because $\displaystyle 
\biggl(\frac{\sigma^{\scriptscriptstyle (-)}}{2}\biggr)^2=0$.
The proof that all the $\widehat{\overline{c}}$ anticommute is the same as the previous one 
with $\sigma^{\scriptscriptstyle (-)}$
replaced everywhere by $\sigma^{\scriptscriptstyle (+)}$. 

The only thing that remains to be proved is the result of the anticommutator of $\widehat{c}$
with $\widehat{\overline{c}}$:
If $k<l$ we have
\begin{equation}
\displaystyle \widehat{c}^k\widehat{\overline{c}}_l=
{\bf 1}\otimes {\bf 1}\otimes\ldots
\otimes {\bf 1}\otimes \underbrace{\frac{\sigma^{\scriptscriptstyle (-)}}{2}\sigma_z}_{k}
\otimes \sigma_z\otimes\sigma_z\otimes
\ldots\otimes \underbrace{\frac{\sigma^{\scriptscriptstyle (+)}}{2}}_{l}\otimes {\bf 1}\otimes\dots\otimes {\bf 1}
\end{equation}
while
\begin{equation}
\displaystyle \widehat{\overline{c}}_l\widehat{c}^k=
{\bf 1}\otimes {\bf 1}\otimes\ldots
\otimes {\bf 1}\otimes \underbrace{\sigma_z\frac{\sigma^{\scriptscriptstyle (-)}}{2}}_{k}
\otimes \sigma_z\otimes\sigma_z\otimes
\ldots\otimes \underbrace{\frac{\sigma^{\scriptscriptstyle (+)}}{2}}_{l}\otimes {\bf 1}\otimes\dots\otimes {\bf 1}.
\end{equation}
Therefore from $\displaystyle \biggl[\sigma_z,\frac{\sigma^{\scriptscriptstyle (-)}}{2}\biggr]_{\scriptscriptstyle +}=0$
we get that $[\widehat{c}^k,\widehat{\overline{c}}_l]_{\scriptscriptstyle +}=0$.
If $k>l$ we have 
\begin{equation}
\displaystyle \widehat{c}^k\widehat{\overline{c}}_l=
{\bf 1}\otimes {\bf 1}\otimes\ldots
\otimes {\bf 1}\otimes \underbrace{\sigma_z\frac{\sigma^{\scriptscriptstyle (+)}}{2}}_{l}
\otimes \sigma_z\otimes\sigma_z\otimes
\ldots\otimes \underbrace{\frac{\sigma^{\scriptscriptstyle (-)}}{2}}_{k}\otimes {\bf 1}\otimes\dots\otimes {\bf 1}
\end{equation}
and 
\begin{equation}
\displaystyle \widehat{\overline{c}}_l\widehat{c}^k=
{\bf 1}\otimes {\bf 1}\otimes\ldots
\otimes {\bf 1}\otimes \underbrace{\frac{\sigma^{\scriptscriptstyle (+)}}{2}\sigma_z}_{l}
\otimes \sigma_z\otimes\sigma_z\otimes
\ldots\otimes \underbrace{\frac{\sigma^{\scriptscriptstyle (-)}}{2}}_{k}\otimes {\bf 1}\otimes\dots\otimes {\bf 1}.
\end{equation}
So from $\displaystyle \biggl[\sigma_z,\frac{\sigma^{\scriptscriptstyle (+)}}{2}\biggr]_{\scriptscriptstyle +}=0$
we have that $[\widehat{c}^k,\widehat{\overline{c}}_l]_{\scriptscriptstyle +}=0$.
Finally if we take the same index $k=l$ we obtain
\begin{eqnarray}
&&\displaystyle \widehat{c}^k\widehat{\overline{c}}_k={\bf 1}\otimes {\bf 1}\otimes \ldots \otimes
{\bf 1}\otimes \underbrace{\frac{\sigma^{\scriptscriptstyle (-)}}{2}\frac{\sigma^{\scriptscriptstyle (+)}}{2}}_{k}\otimes
{\bf 1}\otimes \ldots \otimes {\bf 1}, \nonumber\\
&&\displaystyle \widehat{\overline{c}}_k\widehat{c}^k={\bf 1}\otimes {\bf 1}\otimes \ldots \otimes
{\bf 1}\otimes \underbrace{\frac{\sigma^{\scriptscriptstyle (+)}}{2}\frac{\sigma^{\scriptscriptstyle (-)}}{2}}_{k}\otimes
{\bf 1}\otimes \ldots \otimes {\bf 1}
\end{eqnarray}
from which we can derive
\begin{eqnarray}
\displaystyle [\widehat{c}^k,\widehat{\overline{c}}_k]_{\scriptscriptstyle +}
&=&{\bf 1}\otimes {\bf 1}\otimes \ldots \otimes
{\bf 1}\otimes \underbrace{\biggl[\frac{\sigma^{\scriptscriptstyle (-)}}{2},
\frac{\sigma^{\scriptscriptstyle (+)}}{2}\biggr]_{\scriptscriptstyle +}}_{k}\otimes
{\bf 1}\otimes \ldots \otimes {\bf 1}\nonumber\\
&=&{\bf 1}^{\scriptscriptstyle \otimes 2n}
\end{eqnarray}
where we have used the fact that:
$\displaystyle \biggl[\frac{\sigma^{\scriptscriptstyle (-)}}{2},
\frac{\sigma^{\scriptscriptstyle (+)}}{2}\biggr]_{\scriptscriptstyle +}={\bf 1}.$
So we can conclude by saying that the objects built in (\ref{5-3}) out of the Pauli matrices satisfy the Grassmann algebra.

\section{Appendix }
{\centerline{\bf Calculational Details}

In this Appendix we will give the details of how to construct the representation (\ref{5-10}) of the Grassmannian
variables. An empiric rule which emerges from (\ref{basis}) is the one we shall now illustrate. Let us consider, in the
case $n=1$, the 
reference string $c^pc^q$ and compare it with the four objects $(1,c^p,c^q,c^pc^q)$. For example ``1" has two 
Grassmannian variables {\it lacking} with respect to the reference string, $c^p$ instead has the first
Grassmannian variable {\it present} while the second one is {\it lacking}. The empiric rule we shall use
is that the {\it lacking} of a Grassmannian variable will be indicated by the vector $\left( \begin{array}{c}
1\\ 0 \end{array}\right)$ while the {\it presence} of it by $\left( \begin{array}{c}
0\\ 1 \end{array}\right)$. So this rule gives
\begin{equation}
1\Leftrightarrow \left( \begin{array}{c} 1\\ 0 \end{array}\right)\otimes \left( \begin{array}{c} 1\\ 0 \end{array}\right).
\end{equation}
The $c^p$ has the first {\it present} and the second {\it lacking} so
\begin{equation}
c^p\Leftrightarrow \left( \begin{array}{c} 0\\ 1 \end{array}\right)\otimes 
\left( \begin{array}{c} 1\\ 0 \end{array}\right)
\end{equation}
and so on. If we now apply the same rule for 2 degrees of freedom ($n=2$) considering 
$c^{p_1}c^{q_1}c^{p_2}c^{q_2}$ as reference string
we obtain exactly Eq. (\ref{5-10}). In fact for example $c^{q_2}$ has the
first three variables absent with respect to the reference string and only the last present, so its representation
will be given by 
\begin{equation}
c^{q_2}\Leftrightarrow \left( \begin{array}{c} 1\\ 0 \end{array}\right)\otimes
\left( \begin{array}{c} 1\\ 0 \end{array}\right)\otimes \left( \begin{array}{c} 1\\ 0 \end{array}\right)
\otimes \left( \begin{array}{c} 0\\ 1 \end{array}\right).
\end{equation}

\end{document}